\documentclass[acmtog,nonacm,authorversion=true]{acmart}
\acmSubmissionID{308}
\usepackage{textgreek}

\usepackage[utf8]{inputenc}
\usepackage{booktabs} 
\usepackage{subcaption}
\usepackage{bm}
\usepackage{amsfonts}
\usepackage{booktabs}
\usepackage{mathtools}
\usepackage{nicefrac}
\usepackage{xspace}
\usepackage{algorithm}
\usepackage{algorithmicx}
\usepackage{xcolor}
\usepackage{xspace}
\usepackage{libertine}
\usepackage[shortlabels]{enumitem}
\usepackage[noend]{algpseudocode}
\usepackage{fancyhdr}

\usepackage{tcolorbox}
\usepackage{multirow}
\usepackage[newfloat,frozencache]{minted} 

\usepackage{mdframed}
\usepackage{setspace}
\definecolor{hlcol}{RGB}{255,200,200}
\newminted{cpp}{autogobble,escapeinside=||,fontsize=\footnotesize}
\newminted{python}{autogobble,escapeinside=||,fontsize=\footnotesize}

\BeforeBeginEnvironment{minted}{
    \begin{tcolorbox}[arc=1pt,colback=black!3!white,colframe=black!16!white,boxrule=0.3mm,before skip=1mm,after skip=1mm,top=.5mm,bottom=.5mm,left=1mm,right=1mm]
    }
\AfterEndEnvironment{minted}{\end{tcolorbox}}
\AtBeginEnvironment{minted}{\setstretch{1.2}}

\newcommand{\drjit}{\textsc{Dr.Jit}\xspace}
\newcommand{\mitsuba}{\textsc{Mitsuba 3}\xspace}

\definecolor{darkblue}{rgb}{0,0.0,.6}
\newcommand{\mitsubaurl}{{\color{darkblue}\url{https://github.com/mitsuba-renderer/mitsuba3}}\xspace}
\newcommand{\drjiturl}{{\color{darkblue}\url{https://github.com/mitsuba-renderer/drjit}}\xspace}

\newcommand{\bdelta}{\bm{\delta}}
\newcommand{\vx}{\mathbf{x}}
\newcommand{\vy}{\mathbf{y}}
\newcommand{\mJ}{\mathbf{J}}

\setcitestyle{nosort,square} 

\author{Wenzel Jakob}
\affiliation{
    \institution{École Polytechnique Fédérale de Lausanne (EPFL)}
    \country{Switzerland}
}

\author{Sébastien Speierer}
\affiliation{
    \institution{École Polytechnique Fédérale de Lausanne (EPFL)}
    \country{Switzerland}
}

\author{Nicolas Roussel}
\affiliation{
    \institution{École Polytechnique Fédérale de Lausanne (EPFL)}
    \country{Switzerland}
}

\author{Delio Vicini}
\affiliation{
    \institution{École Polytechnique Fédérale de Lausanne (EPFL)}
    \country{Switzerland}
}

\setcopyright{acmlicensed}\acmJournal{TOG}
\acmYear{2022}\acmVolume{41}\acmNumber{4}\acmArticle{1}\acmMonth{7} \acmDOI{10.1145/3528223.3530099}

\begin{document}
\title{\drjit: A Just-In-Time Compiler for Differentiable Rendering}

\begin{abstract}
    \drjit is a new just-in-time compiler for physically based rendering and
    its derivative. \drjit expedites research on these topics in two ways:
    first, it traces high-level simulation code (e.g., written in Python) and
    aggressively simplifies and specializes the resulting program
    representation, producing data-parallel kernels with state-of-the-art
    performance on CPUs and GPUs.

    Second, it simplifies the development of differentiable rendering
    algorithms. Efficient methods in this area turn the derivative of a
    simulation into a simulation of the derivative. \drjit provides
    fine-grained control over the process of automatic differentiation to help
    with this transformation.

    Specialization is particularly helpful in the context of differentiation,
    since large parts of the simulation ultimately do not influence the
    computed gradients. \drjit tracks data dependencies globally to find and
    remove redundant computation.
\end{abstract}


\begin{teaserfigure}
  \centering
  \includegraphics[width=\linewidth]{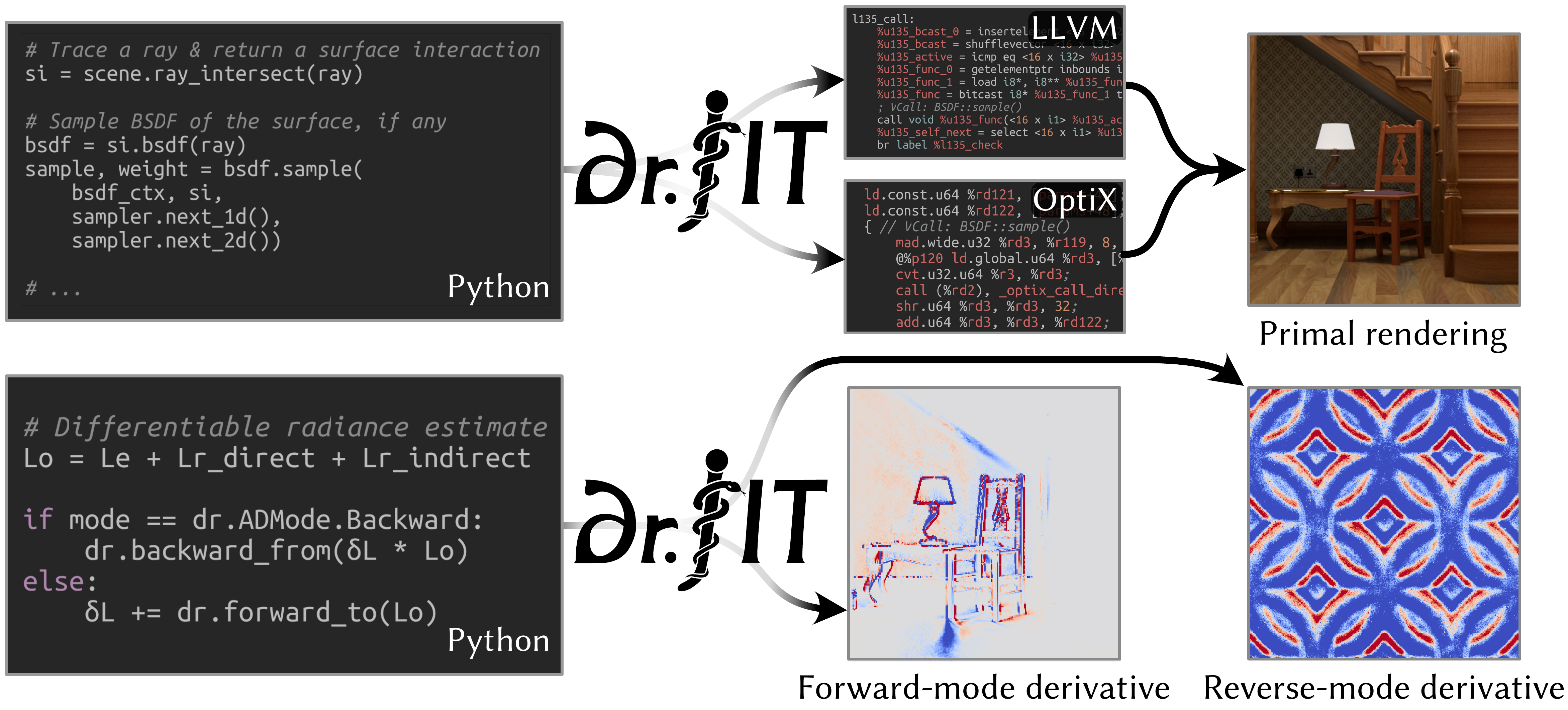}
  \caption{%
    \label{fig:teaser}%
    \drjit is a domain-specific compiler for physically-based (differentiable)
    rendering. When \drjit executes a rendering algorithm, it generates a
    \emph{trace}: a large graph comprised of arithmetic, loops, ray
    tracing operations, and polymorphic calls that exchange information between
    the rendering algorithm and scene objects (shapes, BSDFs, textures,
    emitters, etc.). \drjit specializes this graph to the provided scene and
    compiles it into a large data-parallel kernel (``megakernel'') via LLVM or
    OptiX backends, achieving geometric mean GPU speedups of $3.70\,\times$
    (vs. Mitsuba 2) and $2.14\,\times$ (vs. PBRT 4). While helpful for ordinary
    rendering, the main purpose of \drjit is to dynamically compile
    \emph{differential simulations}. Recent methods in this area decompose a
    larger differentiation task into a series of incremental steps, which
    requires an unusually fine-grained approach to automatic differentiation.
    \drjit supports such transformations in forward and reverse modes: the
    former computes a perturbation in image space, which is helpful for
    debugging and visualization. The latter provides derivatives in parameter
    space (e.g. texels of the wallpaper) for simultaneous optimization of large
    numbers of unknowns.
}
\end{teaserfigure}

\maketitle

\section{Introduction}
\label{sec:intro}
Recent progress in the area of \emph{physically based differentiable rendering}
(henceforth ``PBDR'') has led to the development of methods that can differentiate light
transport simulations with respect to arbitrary scene parameters. 
Combined with a gradient-based optimizer, they can solve nonlinear problems involving large sets of unknowns.
Diverse scientific and engineering disciplines require
the inverse analysis of images and stand to benefit from these developments.

While the theory of PBDR continues to evolve, practical aspects
have remained a persistent challenge. For example, the reverse-mode
derivative of a conceptually simple algorithm like path
tracing~\cite{KajiyaRenderingEquation} with precautions for linear time
complexity~\cite{Vicini2021} and unbiased visibility
handling~\cite{Bangaru2020} turns into an enormously complicated function. 
Correct
implementation of such a large and intricate program is
near-impossible even for experts in the field. 
Mere correctness
is also unsatisfactory: optimizations tend to run for thousands of iterations, hence
the resulting program needs to be fast. It is evident that better tools
are need bridge this conspicuous gap between PBDR theory and practice.

The design of \drjit was guided by a single unifying objective: it
should provide a practical and efficient foundation for work in this area. Most
architectural decisions are direct consequences of this
overarching goal. For example, consider the differentiation step that is
implicit in differentiable rendering. Manual differentiation is tedious and
error-prone, hence it is logical that the system should build on
\emph{automatic differentiation}~(AD) to simplify development.

However, the needs of PBDR are more specific: standard use of AD to
differentiate a rendering algorithm produces an inefficient and biased
derivative that precludes many applications. Recent work addresses
inefficiencies using physical reciprocity~\cite{NimierDavid2020} and arithmetic
invertibility~\cite{Vicini2021} to turn the derivative of a simulation into a
simulation of the derivative, while re-parameterizing the integration domain to
remove bias~\cite{Loubet2019}. These steps move the differentiation operation
into the random walk, where it introduces partial derivative terms at each
scattering event. This has implications on the design of the system: the
derivatives must somehow be (pre-)compiled, since the machinery of AD is too
slow to be \mbox{used dynamically at such high rates.}

Differentiating a simulation changes the underlying computation, but the
details of this change depend on the scene, simulation algorithm, and
optimization task. When the optimization only targets a subset of the scene's
parameters, it is desirable that the system uses this information to remove
steps that cannot influence the computed gradient. The dynamic nature of this
problem calls for a similarly dynamic approach to compilation, which is why we
pursue an approach centered around \emph{just-in-time} (JIT) compilation.

Effective use of modern computing hardware requires that the program is
organized into data-parallel phases known as \emph{kernels}. Several kernels
are generally needed to handle data dependencies, which must then exchange
information through device memory. This inter-kernel communication
comes at a cost in terms of storage and memory bandwidth, hence the specific manner in which a
computation is partitioned into kernels can have a pronounced impact on
performance. In the case of PBDR, the simulation parallelizes over
millions of Monte Carlo samples that represent a large amount of program
state. In our experiments, we find that it is almost always preferable that the
Monte Carlo integration occurs within a \mbox{\emph{megakernel}}, i.e.,  a
large kernel containing all program instructions needed to evaluate the
integrand. Most sample state can then be stored in
registers, reducing memory usage \mbox{and inter-kernel communication.}

Finally, physically-based rendering algorithms are commonly expressed using
\emph{subtype polymorphism} to dynamically dispatch method calls from abstract
component interfaces (e.g., a material encountered by a ray) to concrete
implementations (e.g., a woven fabric or a rough metallic surface). The ability to
represent, differentiate, and optimize such polymorphic constructions \mbox{benefits
performance.}

Taking stock, we arrive at the following set of requirements:
\begin{enumerate}[leftmargin=6.5mm,label=\arabic*.]
    \item The system must dynamically generate specialized code
        for a given scene, rendering algorithm, and optimization task.
        \\[-3.0mm]

    \item Compilation should be able to ensure that any use of Monte Carlo
        integration remains fully contained within a megakernel.
        \\[-3.0mm]

    \item The system must scale. Challenges include fine-grained AD,
        thousands of volume scattering events, large numbers of shapes and
        materials accessed \mbox{through polymorphic abstractions.}
\end{enumerate}

\drjit addresses these requirements using an approach based on \emph{tracing}.
It executes simulation code in a deferred manner by recording encountered
operations into a \emph{trace} for subsequent compilation and execution on a
target device. This process is automatic and language-agnostic (Python and C++
are currently supported).

A key difference to Mitsuba 2~\cite{NimierDavid2019}, which also uses tracing,
is that \drjit must handle a larger set of operations to guarantee successful
megakernel generation. A \drjit (differential) rendering step captures loops,
polymorphism, and ray intersection operations without interruption, returning
an unevaluated image in the form of a \emph{very large} trace that encompasses
the rendering algorithm and implementations of all referenced scene objects
including materials, textures, volumes, light sources, etc.

This global representation reveals optimization opportunities. For example,
suppose that rendering does not integrate over time: it is then safe to remove
time-related variables from loops and function interfaces. Similarly, the
derivatives of many program variables do not influence the computed parameter
gradient and can be deleted.

\drjit finally compiles and evaluates the trace via OptiX~\cite{OptiX} or
LLVM~\cite{LLVM:CGO04}. The former produces GPU kernels leveraging ray tracing
hardware acceleration, while the latter generates vectorized code \mbox{for
diverse CPU architectures.}

In addition to \drjit, we also present \mitsuba, a new version of the Mitsuba
renderer that builds on \drjit. Both projects are available under an open
source license at \drjiturl and \mitsubaurl.

Differentiable rendering subsumes ordinary rendering, and \drjit achieves
state-of-the-art performance on both tasks. We substantiate all
performance-related observations experimentally. Following a review of related
work, the remainder of this article discusses compilation
(Section~\ref{sec:system-jit}) and differentiation
\mbox{(Section~\ref{sec:system-ad}) in turn.}

\section{Related work and Background}
\label{sec:related}

\subsection{Array programming}
Increasing interest in machine learning in recent years has precipitated the
creation of numerous frameworks that combine AD with $n$-dimensional array
representations. They use JIT backends like XLA~\cite{XLA} to fuse operations
into efficient kernels. Given widespread success and obvious similarities to
\drjit, it is not unreasonable to wonder whether this article could have been
cut short by a recommendation to implement PBDR methods on top of such a
framework? We investigate this question in Section~\ref{sec:system-jit} and
find that PBDR workloads exhibit characteristics that are unusual in the array
programming setting, causing them to fall off the fast path.

\subsection{Automatic Differentiation}

Manual differentiation of mathematical expressions is error-prone and
mechanical. The natural desire to delegate this task to a computer began with
pioneering work in the 1950s and
1970s~\cite{nolan1953analytical,Wengert1964,Linnainmaa1976} followed by
comprehensive study in the
1980s~\cite{speelpenning1980compiling,griewank1989automatic}. Griewank and
Walther's book~\shortcite{Griewank2008} reviews what has been learned about AD
over the course of these many decades.

Given the consolidated understanding of the mathematical structure and
asymptotic complexity of various derivative propagation strategies, there is a
surprising degree of variety when it comes to how AD should be exposed to the
user. The space of methods includes tracing, source-to-source transformation of
abstract syntax trees or \emph{intermediate representations} (IR), and hybrids
combining tracing with transformation. Differentiation can target scalar,
dense, or sparse array programs in forward, reverse-, and mixed modes,
computing gradients or higher-order derivatives of pure and impure functions.
Covering all techniques in detail is far beyond the scope of this paper, and we
refer to a recent survey by Baidin et al.~\shortcite{Baydin2018Automatic}. This
section only covers core concepts, noteworthy related methods for first-order
\mbox{derivatives, and their relationship to our approach.}

\paragraph{Directionality.} 

The main high-level flavors of AD are the \emph{forward} and \emph{reverse}
(also known as \emph{adjoint} or \emph{backward}) modes. Forward mode evaluates
a \emph{Jacobian-vector-product} (often abbreviated ``JVP'') of the form
$\bdelta\vy=\mJ_f\bdelta\vx$, where $\mJ_f$ is the Jacobian of the
\emph{primal} (i.e., original) computation $\mathbf{y}=f(\mathbf{x})$, and
reverse mode evaluates a \emph{vector-Jacobian-product} (``VJP'') of the form
$\bdelta\vx=\bdelta\vy^T\mJ_f$. Both can in principle compute the same
derivatives, but forward mode does this more efficiently when the function
being differentiated has few inputs (ideally just one), while reverse mode is
efficient if it has few outputs. Realistic scene descriptions have million of
unknowns, hence practical differentiable rendering depends on reverse mode.

\paragraph{Reverse mode.}
The key issue with reverse mode is that it inverts the data dependencies of the
original program. The derivative of this reversed program references
intermediate steps of the primal calculation, which raises the age-old question
of how they should be obtained. Exhaustive storage is simple but does not
scale, as modern processors can generate many terabytes of intermediate state
per second. The usual remedy is to only store this state at a sparse set of
\emph{checkpoints} with later recovery via reevaluation from the nearest
one~\cite{Volin1985automatic}. Automatic recursive usage of this pattern
reduces storage and runtime overheads of a program with $t$ \mbox{operations to
a factor of $O(\log t)$~\cite{Siskind2018divide}.}

\paragraph{Adjoints.} 
When available, \emph{custom adjoints} are generally preferable to
checkpointing. These differentiation techniques exploit problem-specific traits
to reduce storage and reevaluation overheads. For example, differentiating
through all steps of a multivariate Newton's method is unnecessarily
inefficient, since the implicit function theorem provides the answer directly
from the iteration's fixed point. Similarly, the solution of an ordinary
differential equation admits an efficient adjoint that \mbox{reverses
time~\cite{pontryagin1962}}.
\addtolength{\textfloatsep}{-4mm}
\begin{figure}[b]
\begin{minipage}{4.0cm}
\begin{minted}[fontsize=\footnotesize]{python}
def pow(x: float, n: int):
    y = 1
    for i in range(n):
        y *= x
    return y
\end{minted}
\begin{minted}[fontsize=\footnotesize]{python}
def pow_jvp(x, n, dx):
    y, dy = 1, 0
    for i in range(n):
        dy = x*dy + y*dx
        y *= x
    return dy
\end{minted}
\end{minipage}
\begin{minipage}{4.4cm}
\begin{minted}[fontsize=\footnotesize]{python}
def pow_vjp(x, n, dy):
    y, dx = 1, 0
    stack = []
    for i in range(n):
        stack.append(y)
        y *= x
    for i in reversed(range(n)):
        y = stack.pop()
        dx += y*dy
        dy *= x
    return dx
\end{minted}
    \vspace{2.7mm}
\end{minipage}
\vspace{-3mm}
\caption{%
    \label{fig:tapenade-example}%
    Derivatives of a program obtained using \emph{source transformation}. Here, \texttt{pow}
    raises $x$ to the $n$-th power ($n\in\mathbb{N}$). The JVP/VJP versions
    were generated by Tapenade~\cite{Hascoet2013} and translated to
    Python. Both require the primal function arguments as input. The JVP further
    takes the function input derivatives and converts them into output
    derivatives, while the VJP takes output derivatives and converts them into
    input derivatives.
}
\end{figure}

\paragraph{Adjoints for rendering.} 
Light transport also admits custom adjoints: \emph{Radiative
Backpropagation}~(RB)~\cite{NimierDavid2020} and \emph{Path Replay
Backpropagation}~(PRB)~\cite{Vicini2021} transform the derivative of a
simulation into an equivalent and more efficient simulation of \emph{derivative
radiation}. \mbox{Section~\ref{sec:pbdr-review} discusses them further.}

\paragraph{Tracing.} 
Assuming that efficient adjoints are available, the next important question is
how the computation to be differentiated should be ingested. Methods based on
\emph{tracing} record an evaluation trace (also referred to as a \emph{Wengert
tape} or \emph{computation graph}) of all differentiable arithmetic operations;
differentiation traverses this trace once more to propagate derivatives in
forward or reverse mode. The trace ignores control flow constructs and stores
unrolled versions of all loops and taken branches. Tracing has seen widespread
adoption through tools like PyTorch~\cite{PyTorch}, in which a typical neural
network produces a trace with a few hundred high-level operations. The
reverse-mode sweep then invokes efficient adjoints of each operation. Tracing
inherits the usual caveats of reverse mode---for example, differentiating
long-running loops can be challenging. It also adds runtime overheads that can
normally be amortized when differentiating array programs, but they dominate in
\emph{scalar programs} that manipulate individual floating point values. In
iterative computations, tracing is often performed repeatedly, which causes
further overheads and inhibits optimization.

\label{sec:source-trafo}
\paragraph{Source transformation.} 
AD via \emph{source transformation} converts the source code of a program into
its derivative. In essence, differentiation becomes a one-time compilation
step, which enables practical differentiation of scalar programs.
Figure~\ref{fig:tapenade-example} illustrates this on~a~(not~particularly good)
implementation of an integer power $x^n$, with JVP/VJP variants generated by
Tapenade~\cite{Hascoet2013}. A key advantage of this approach is the
preservation of control flow: unlike an unrolled trace, the programs in
Figure~\ref{fig:tapenade-example} work regardless of the value of $n$ (the
number of loop iterations).

Loops continue to be a nuisance, however: observe how the VJP requires two of
them: the first records all intermediate loop state into a \texttt{stack}
variable that is later consumed by a reversed loop. This poses difficulties on
massively parallel architecture like GPUs: stack or \emph{shadow memory} must
be provisioned for all threads running in parallel, while handling worst-case
requirements that are generally unknown. The JVP and VJP are equivalent in this
specific example, since \texttt{pow} only has a single differentiable argument
and return value. 

Curiously, optimizations that are straightforward in one AD approach can become
relatively difficult in another. For example, only a subset of program
variables usually affects the final gradient; it is desirable that the AD
system recognizes this to avoid unnecessary adjoint evaluations. In tracing AD,
this optimization happens automatically, since traversal along data
dependencies cannot reach irrelevant variables. To achieve the same goal,
source transformation tools must perform a more involved \emph{activity
analysis}~\cite{bischof1992adifor}, which is a data-flow analysis that
conservatively propagates derivative liveness through the control flow graph
until this process reaches a fixed point.

Four notable source transformation tools are Tapenade~\cite{Hascoet2013},
Stalin$\nabla$~\cite{Pearlmutter2008reverse}, Zygote~\cite{Innes2019dont}, and
Enzyme~\cite{moses2020instead,moses2021reverse}. Tapenade uses sophisticated
data-flow analyses to generate optimized derivatives of scalar C or Fortran
programs involving pointers and array mutation. Stalin$\nabla$ operates on
$\lambda$-calculus IR and performs source-to-source transformations using
runtime reflection. The \emph{callee-derives} approach in their work shares
some of the motivation of \drjit's specialization of polymorphic derivatives.
Zygote operates on typed IR of the Julia language and composes higher-level
adjoints, while Enzyme differentiates code following lowering to LLVM IR. As
with optimizing compilers, the IR's abstraction level can facilitate or
exacerbate certain tasks, hence the suitability of these tools depends nature
of the problem to be differentiated.

\paragraph{Hybrid systems.}
Tracing and source transformation are merely the extreme points of a large
space of \emph{hybrid} techniques with interesting trade-offs. For example,
PyTorch~\shortcite{PyTorch} programs can be traced into a domain-specific
language for subsequent compilation, which removes tracing overheads and
enables optimizations. JAX~\cite{Jax} traces functions into an expressive IR
that can be differentiated or parallelized. Compilation proceeds via lowering
into XLA~\cite{XLA} that performs further tensorial optimizations. \drjit is
also a hybrid in this classification: it ingests computation using tracing but
generates kernels that preserve the control flow (loops, subroutines) of the
original program. Its output thus resembles that of a source
transformation--based tool.

\paragraph{Functional languages.}
Following Elliot's~\shortcite{SimpleEssence} formalization of backpropagation
using category theory and continuation passing, multiple works have proposed
languages that combine the expressiveness of higher-order functional
programming with the efficiency of imperative languages. This includes the
previously mentioned Stalin$\nabla$~\cite{Pearlmutter2008reverse} and languages
like $\mathrm{d}\tilde{\mathrm{F}}$~\cite{FSharp} and Dex~\cite{Dex}. \drjit
adopts an imperative approach to interoperate with C++ and Python. That said,
the authors find combinations of PBDR with differentiable functional
programming promising and worthy of future exploration.

\subsection{Graphics and compilers}
Graphics applications have a near-insatiable thirst for floating point
operations, usually operating at the limits of what is possible on present
hardware. Their needs are often not well-served by existing programming
languages and compilers, which has motivated specialized systems with broad
impact. For example, the \emph{RSL} shading
language~\cite{hanrahan1990language} pioneered the use of programmability in a
previously mostly static graphics stack, using JIT-compilation and
scene-specific specialization analogous to similar steps in \drjit. Following
increased programmability of GPUs through shading languages like
\emph{Cg}~\cite{Cg}, the \emph{Brook}~\cite{buck2004brook} project was
instrumental in repurposing GPUs for general-purpose computation (``GPGPU'').

\emph{SMASH}~\cite{ShaderMetaprogramming} introduced the idea of
metaprogramming shaders by tracing arithmetic in a host language, and
\emph{Sh}~\cite{ShaderAlgebra} further specialized generated code.
Spark~\cite{Foley2011Spark} raised the level of abstraction by disentangling
features from the underlying GPU pipeline stages, and Slang~\cite{he2018slang}
modularized shader development by separating aspects like lighting, materials,
and camera transformations. The system parses high-level C++-style code and
specializes the resulting IR before emitting low-level shader code for various
platforms.

\begin{figure*}[t]
    \centering
    \includegraphics[width=0.8\textwidth]{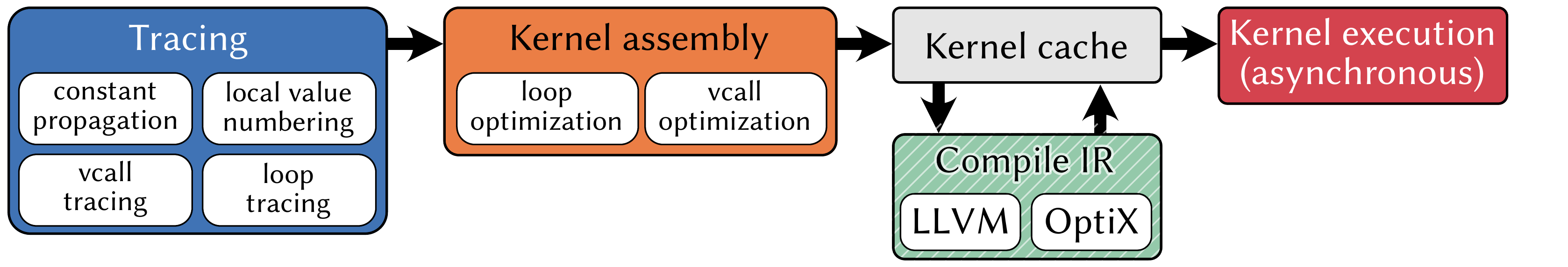}
    \vspace{-3mm}
    \caption{%
        \label{fig:drjit-pipeline}%
        The five main phases of \drjit. \emph{Tracing} executes a Python or C++
        program using custom arithmetic types that record operations into a
        graph data structure. Basic optimizations remove redundancies and
        reduce the size of the program. Tracing of loops and polymorphic
        constructs (\emph{vcalls}, i.e., virtual method calls) requires special precautions at this
        stage. \emph{Kernel assembly} removes redundancies at a global level
        and produces a program in the desired intermediate representation (LLVM
        IR or PTX). Tracing and kernel assembly are highly optimized (on the
        order of 1-15ms in typical cases). A subsequent backend compilation step
        converts the generated IR into executable machine code. Backend
        compilation is relatively costly, hence \drjit consults an
        in-memory and on-disk kernel cache to see if this computation was
        previously encountered. Caching is a good fit for the repetitive
        computation performed by gradient-based optimizers.
    }
\end{figure*}

\emph{Halide}~\cite{Halide} simplifies the design of image processing pipelines
by separating computation and \emph{schedule}, which encompasses placement,
parallelization, vectorization, and blocking. A differentiable
extension~\cite{DiffHalide} enables end-to-end optimization of image processing
pipelines. The predictable structure of the underlying stencils enables highly
effective optimizations like scatter-gather conversion during differentiation.
\emph{Taichi}~\cite{Taichi} is a general-purpose parallel programming language
with an emphasis on dynamical simulation and sparse data structures.
\emph{DiffTaichi}~\cite{Difftaichi} endows \emph{Taichi} with an efficient
differentiation operator that preserves the megakernel structure of the
simulation code. While \emph{DiffTaichi}, \emph{Halide}, and \drjit each target
different types of programs, they share a common focus on optimizing memory
access patterns when a computation is differentiated.

A number of PBR systems rely on specialized compilation techniques:
\emph{MoonRay}~\cite{lee2017vectorized} traces vectorized wavefronts, and
\mbox{\emph{Manuka}~\cite{fascione2018manuka}} JIT-compiles shader graphs for a
vectorized parallel batch shading phase. Building on the
\emph{AnyDSL}~\cite{leissa2018anydsl} partial evaluation framework,
\emph{Rodent}~\cite{Rodent} compiles specialized renders for each scene,
sharing some of the motivation of \drjit.
\emph{Mitsuba~2}~\cite{NimierDavid2019} builds on the \emph{Enoki}~\cite{Enoki}
library to retarget generic specifications of rendering algorithms and scene
objects to diverse applications including differentiation, vectorization,
spectral rendering, and polarization. \drjit was designed as a replacement for
Enoki, enabling experimentation with compilation and differentiation using the
infrastructure of an existing renderer.

\enlargethispage{5mm}
Given the favorable discussion of megakernels in Section~\ref{sec:intro} we
should also review their disadvantages, which were studied by Laine et
al.~\shortcite{laine2013megakernels}: megakernels tend to contain large amounts
of code connected via branch instructions. Branch divergence can then reduce
the effectiveness of vectorized program execution, while high register usage
interferes with the latency-hiding mechanism~of~GPUs. \drjit can be used to
explore the impact of kernel size
{\parfillskip=0pt\par}\noindent
on performance: besides producing megakernels, it is also able to emit
different granularities of \emph{wavefronts}, which refers to a sequence of
smaller kernels with intense data exchange through device memory. Laine et
al.~\shortcite{laine2013megakernels} compared both extremes and found wavefront
execution to be the superior execution model. However, nearly a decade and four
hardware generations later, our experiments indicate that the overall balance
seems to have shifted back towards megakernels, at least for the types of
workloads investigated in this article. We expect that there will still be a
point where a megakernel is simply \emph{too large} to run reasonably on a GPU.
\addtolength{\textfloatsep}{+5mm}

\vspace{-1mm}
\section{Just-In-Time Compilation}
\label{sec:system-jit}

We now turn to \drjit's computational substrate and postpone most discussion of
differentiation to Section~\ref{sec:system-ad}. Any use of the term
\emph{tracing} in this section thus refers to capturing computation for later
compilation and is unrelated to AD. The term is also not to be confused with
\emph{ray tracing}, which refers to a specific geometric intersection operation
that can appear within a trace.

Figure~\ref{fig:drjit-pipeline} illustrates the high-level pipeline: following
tracing, kernel assembly generates IR requiring
backend compilation into machine code. In typical PBDR usage, this is only
necessary during the first gradient descent step. The final pipeline stage
launches the kernel for asynchronous execution on the CPU (via a thread pool)
or the GPU (via \emph{CUDA}/\emph{OptiX}) so that tracing and kernel execution
can continue in parallel. \emph{OptiX}~\cite{OptiX} is a domain-specific
compiler that accelerates ray tracing on NVIDIA GPUs, and which offloads these
operations onto dedicated hardware cores if available.

\vspace{-1mm}
\subsection{Running example}

We will now walk through the implementation of a simple \emph{ambient
occlusion} integrator and use this as an opportunity to introduce major
system components, starting with low-level details and then progressively
zooming out. Our program begins by importing\footnote{The shown source code
fragments use the Python language, but the system is also usable via a
near-identical C++ API. JIT-compilation and differentiation usually trace
combinations of code written in both languages.
Operations like \texttt{linspace} and \texttt{meshgrid}
imitate their eponymous counterparts in other array programming
tools.} \drjit along with a floating
point and an integer type.
\begin{minted}[fontsize=\small]{python}
import drjit as dr
from drjit.cuda import Float, UInt32
\end{minted}
\noindent While types like \texttt{Float} and \texttt{UInt32} are suggestive of an
internal scalar representation, they represent dynamically sized 1D arrays.
Instances of them are best thought of as a scalar variable declaration within a loop of
the form ``\mintinline{Lua}{for index in range(..)}'', where \texttt{index}
will usually refer to the Monte Carlo sample being computed.

Operations involving these capitalized types become part of the trace, while
builtin Python types (\mintinline{Lua}{float}, \mintinline{Lua}{int}) and
control flow statements are invisible to \drjit. In other words, this is a
\emph{metaprogram}, whose execution determines what computation will eventually
take place on the target device.

The next line creates a \texttt{Float} variable containing 1024 evenly spaced
numbers covering the interval $[-1, 1]$ that we will shortly use to generate
primary camera rays.
\vspace{.5mm}
\begin{minted}[fontsize=\small]{python}
x = dr.linspace(Float, -1, 1, num=1024)
\end{minted}
\vspace{.5mm}
\noindent The trace of this expression consists of five variables:
the aforementioned loop \texttt{index} variable, an
int-to-float cast, and a \emph{fused multiply-add} (FMA) referencing two
\emph{literal constants} shown in green:
\begin{center}
    \vspace{-.5mm}
    \includegraphics[width=.8\columnwidth]{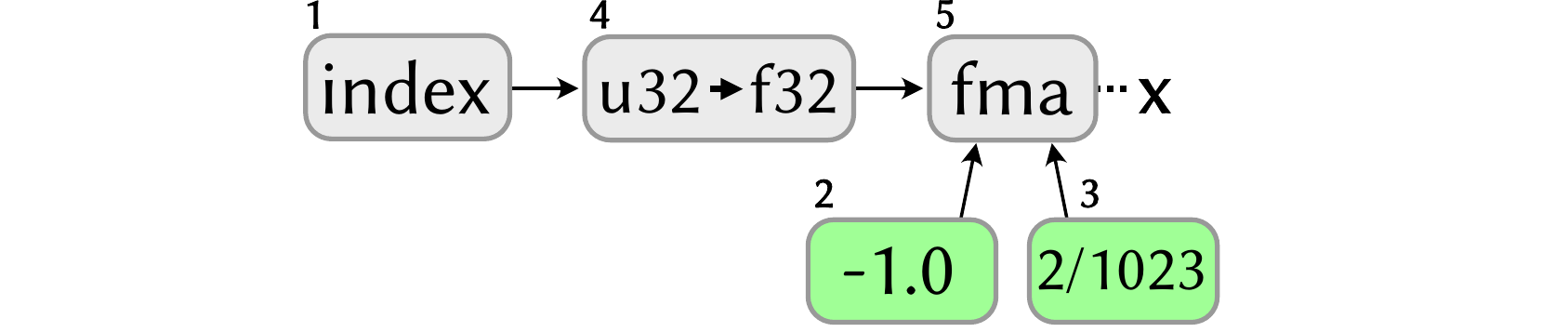}
    \vspace{-.5mm}
\end{center}
A variable trace usually contains at least one use of the \texttt{index}
variable, otherwise it is uniform. Each temporary/variable is identified by
a number ($\texttt{x}$, e.g., points to \#5). An associative data structure maps this
index to a record describing the operation and its dependencies. The next step
of our example creates another 1D array and invokes \texttt{dr.meshgrid} to
expand both arrays \mbox{into 2D grid coordinates.}
\vspace{.5mm}
\begin{minted}[fontsize=\small]{python}
y = dr.linspace(Float, -1, 1, num=1024)
x, y = dr.meshgrid(x, y)
\end{minted}
\vspace{.5mm}
\noindent Several points are worthy of note: the computation of the \texttt{y}
variable is of course redundant. \drjit detects this using
\emph{local value numbering}, an optimization that uses the variable
details (operation type, input dependencies) as \emph{key} to query
an auxiliary \emph{inverse} version of the associative mapping. If
an equivalent variable exists, it will be reused.
The system also performs constant
folding/propagation and basic algebraic simplifications (e.g. \texttt{fma(a,
b, 0) = a*b}; \texttt{a*1=a}) while tracing. These optimizations are not
important in our example, but they are effective in combination with AD
that tends to generate many operations of this type. Both steps are
cheap to do while tracing, and they reduce the size of the IR
passed to OptiX/LLVM.

The function \texttt{dr.meshgrid()} computes row and column indices and then
calls \texttt{dr.gather()} to read from the input arrays \texttt{x} and
\texttt{y}. Both represent computation that has not occurred yet, and
\texttt{dr.gather} is thus able to perform the indexing operation symbolically
by cloning their graph representation and rewriting the \texttt{index}
variable. Along with further application of the previously mentioned
optimizations (value numbering, constant propagation), this produces the
following combined trace:
\begin{center}
    \includegraphics[width=.8\columnwidth]{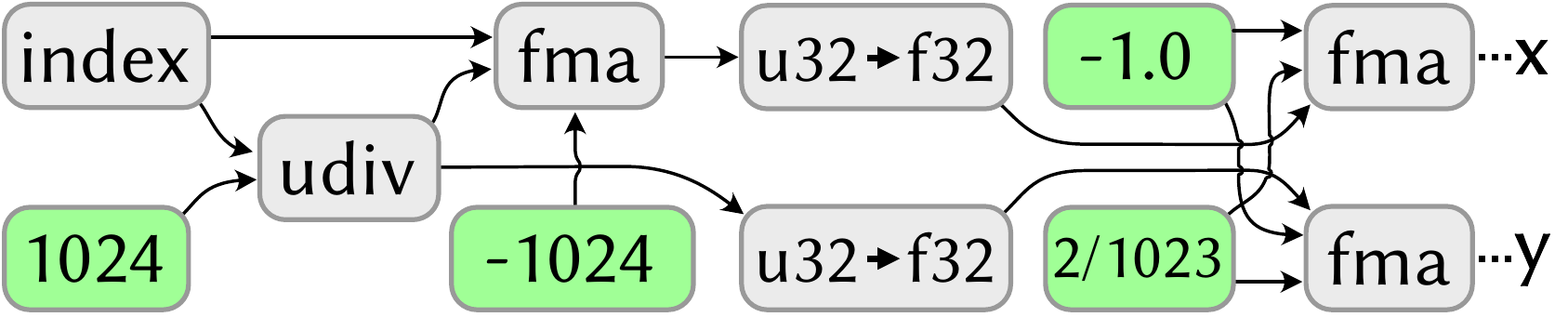}
\end{center}
\noindent Suppose that the example takes place in an interactive session, and
the user wishes to check the contents of the \texttt{x} variable at this point.
\begin{minted}[fontsize=\footnotesize]{python}
>>> print(x)
[-1.0, -0.998, -0.996, .. 1048570 skipped .., 0.996, 0.998, 1.0]
\end{minted}
\noindent Accessing array contents triggers evaluation via
\texttt{dr.eval(..)}, which compiles and launches a fused kernel that commits the
requested variables(s) to device memory. Further use of \texttt{x}
references the stored version, \mbox{hence parts of the trace that are no longer needed
expire.}
\begin{center}
    \includegraphics[width=.8\columnwidth]{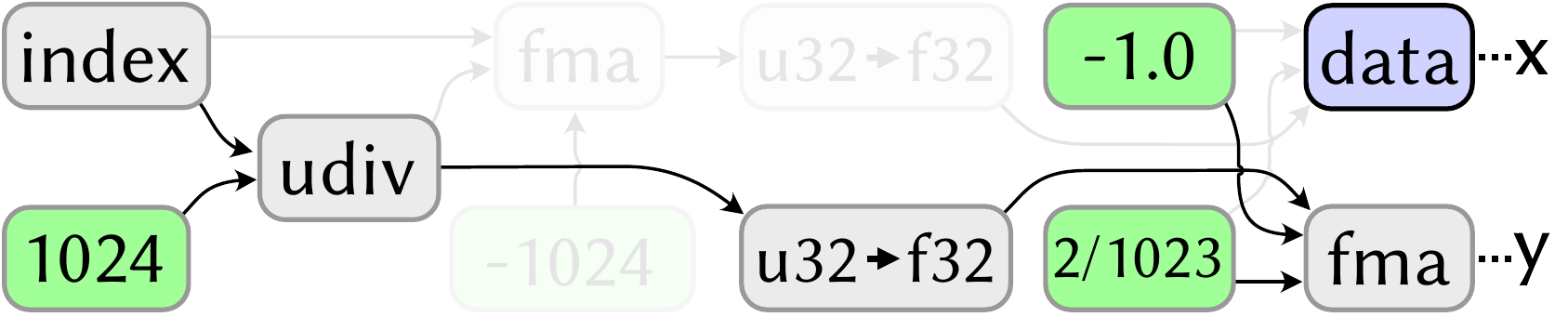}
\end{center}
We are now almost ready to process a set of primary rays, using the 2D grid as
image-space offsets in an orthographic camera model:
\vspace{.8mm}
\begin{minted}[fontsize=\footnotesize]{python}
from mitsuba import Ray3f, Point3f, Vector3f, load_file
ray = Ray3f(o=Point3f(x, y, 0), d=Vector3f(0, 0, 1))
scene = load_file('scene.xml')
si = scene.ray_intersect(ray)
\end{minted}
\vspace{.8mm}
\noindent \mitsuba uses \drjit-provided types
throughout all interfaces; fixed-size arrays like \texttt{Point3f}
simply wrap several JIT variables that become part of the generated program.
The ray tracing operation requires further elaboration: its implementation
looks as follows:
\vspace{.8mm}
\begin{minted}[fontsize=\footnotesize]{python}
class Scene:  # .. (most definitions omitted) ..
    def ray_intersect(self, ray: Ray3f) -> SurfaceInteraction3f:
        pi = self.ray_intersect_preliminary(ray)
        return pi.shape.compute_surface_interaction(pi)
\end{minted}
\vspace{.8mm}
\noindent It first finds a \emph{preliminary intersection} comprised of the
intersected shape, primitive ID, ray depth and UV coordinates (our example is
simplified here to hide complexities like instancing). The second step invokes
a method on the intersected shape (\texttt{pi.shape}) so that it can refine
this preliminary
intersection into a detailed \texttt{SurfaceInteraction3f} type describing the
differential geometry at the intersection point. Within \mitsuba, this
refined intersection maps
to a large number of variables (41-45 depending on the variant of the
renderer). Following these last steps, the trace looks as follows (repeated
parts on the left are omitted):
\begin{center}
    \includegraphics[width=.8\columnwidth]{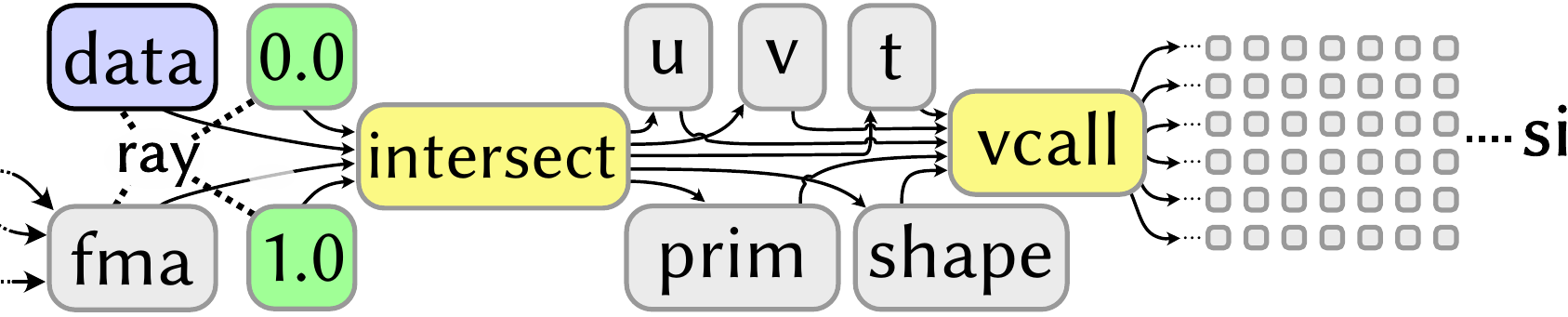}
\end{center}
The yellow boxes represent complex operations with custom code generation
hooks. During kernel assembly, \texttt{intersect} produces
IR that invokes a ray tracing accelerator (Embree~\cite{Embree} on
the CPU and OptiX~\cite{OptiX} on the GPU).
\begin{figure*}[t]
    \centering
    \includegraphics[width=.82\textwidth]{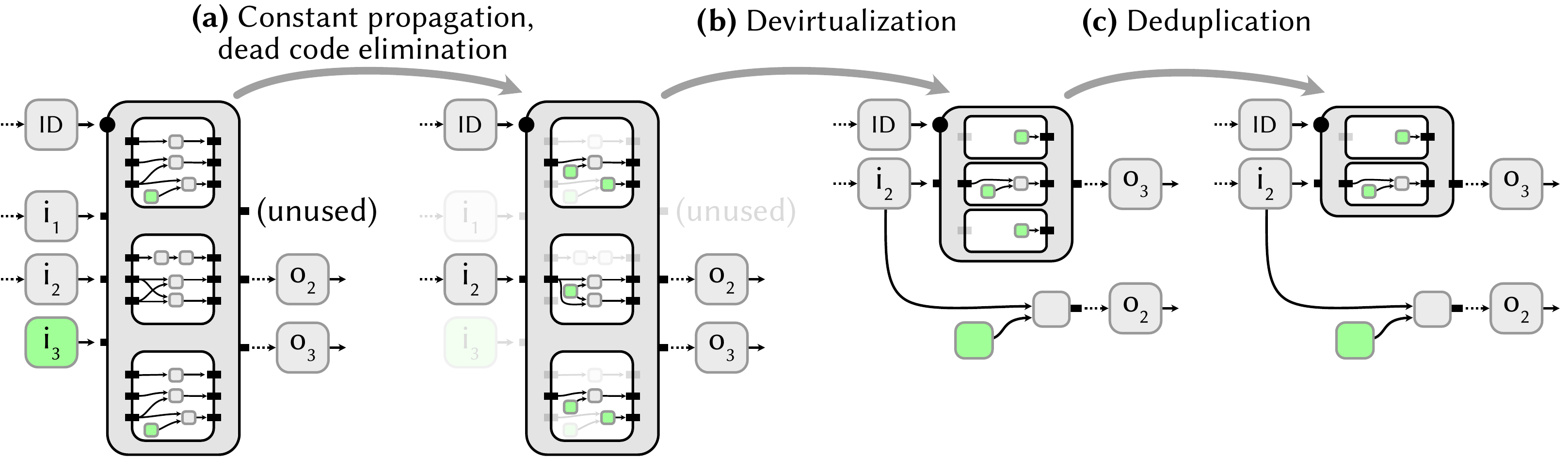}
    \vspace{-2.5mm}
    \caption{%
    \label{fig:vcall-opt}%
    When \drjit encounters a polymorphic method call, it traces the
    implementation of all reachable instances and performs multiple
    optimizations. In the example shown above, a call with inputs $(i_1, i_2,
    i_3)$ returns outputs $(o_1, o_2, o_3)$. The variable $i_3$ is a constant
    literal, and the surrounding program only references outputs $o_2$ and
    $o_3$. \textbf{(a)}. \drjit propagates the constant into the captured
    sub-traces, while at the same time eliminating dead code across the call
    boundary. \textbf{(b)}. In this example, all sub-traces perform the same
    computation to produce $o_2$, and the computation is subsequently
    \emph{devirtualized}, i.e., moved out of the call. \textbf{(c)}. Finally,
    two of the sub-traces are found to be identical and only produce a single
    function definition during kernel assembly.}
\end{figure*}

The \texttt{pi.shape} member of the preliminary intersection refers to an
(as of yet unevaluated) array of $1024\times 1024$ pointers to arbitrary shape
implementations. The yellow operation labeled ``\texttt{vcall}'' represents the dynamic dispatch
step (\emph{virtual function call} in C++ terminology) needed to
resolve the polymorphic nature of this operation. \drjit does so by tracing
\emph{all reachable} method implementations (scene objects register themselves
with \drjit to enable this). The operation can thus be interpreted
as a large-scale demultiplexer-multiplexer that routes arguments and return
values to/from instances. During kernel assembly, it produces IR performing an
indirect branch to one of multiple subroutines (in other words, the indirection
is preserved in the generated program).
\begin{center}
    \includegraphics[width=.8\columnwidth]{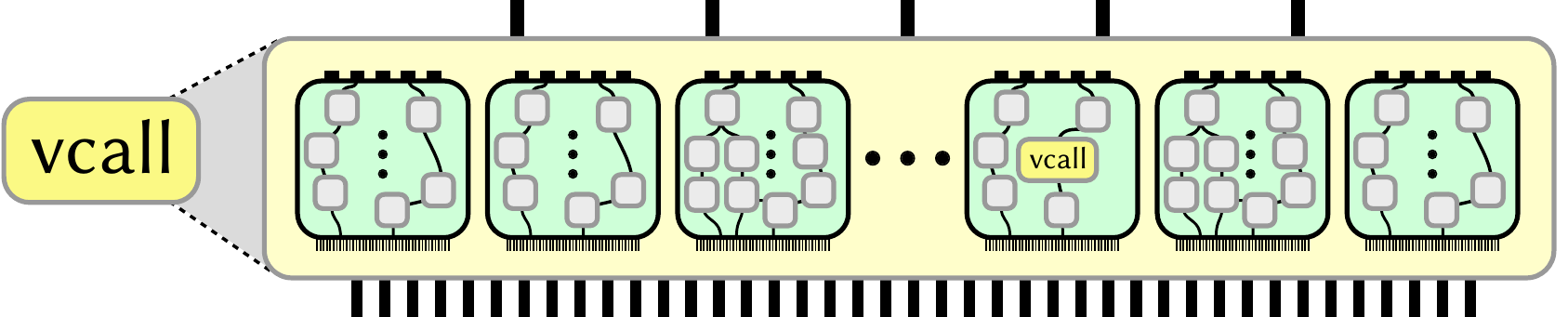}
\end{center}
Several inefficiencies may be apparent to the reader at this point; we will
soon explain how they are resolved. The final part of our example
traces \texttt{128} ambient occlusion rays starting at the
intersected position. For this, we require a random number generator and a
looping mechanism. While a standard Python \mintinline{python}{for} loop works,
it would unroll the loop body 128 times and produce an unnecessarily large
trace. As before, it is desirable that tracing preserves the structure of the
input program. We instead instantiate a \drjit \texttt{Loop} object along with
a pseudorandom number generator (\texttt{PCG32}), a counter (\texttt{i}),
and a variable representing the final result.\nocite{PCG32}
\vspace{.8mm}
\begin{minted}[fontsize=\footnotesize]{python}
from drjit.cuda import Loop, PCG32
rng = PCG32(size=1024 * 1024)
i, result = UInt32(0), Float(0)
loop = Loop(state=lambda: (rng, i, result))
while loop(si.is_valid() & (i < 128)):
    # ... loop body ...
    i += 1
\end{minted}
\vspace{.8mm}
\noindent
This loop runs for only \emph{one} iteration (the condition \mintinline{python}{loop(..)} returns
\mintinline{Python}{False} in the second round). This suffices to capture the
effects of an individual loop iteration, which is all that is needed to
wire\footnote{This involves the insertion of basic block boundaries and \texttt{Phi} nodes into the generated IR,
which is in static single assignment (SSA) form.} it into the generated kernel.
A downside of our approach for embedding tracing into a host language is that
\drjit must know about the loop's \emph{state variables}, which refers to
variables that are modified in the body and either accessed in subsequent
iteration or
following termination. They are specified using a lambda function, which our
implementation invokes before and after the loop to catch redefinitions of
Python variables.
The trace now looks as follows (\texttt{rng} seeding omitted
for simplicity).
\begin{center}
    \includegraphics[width=.8\columnwidth]{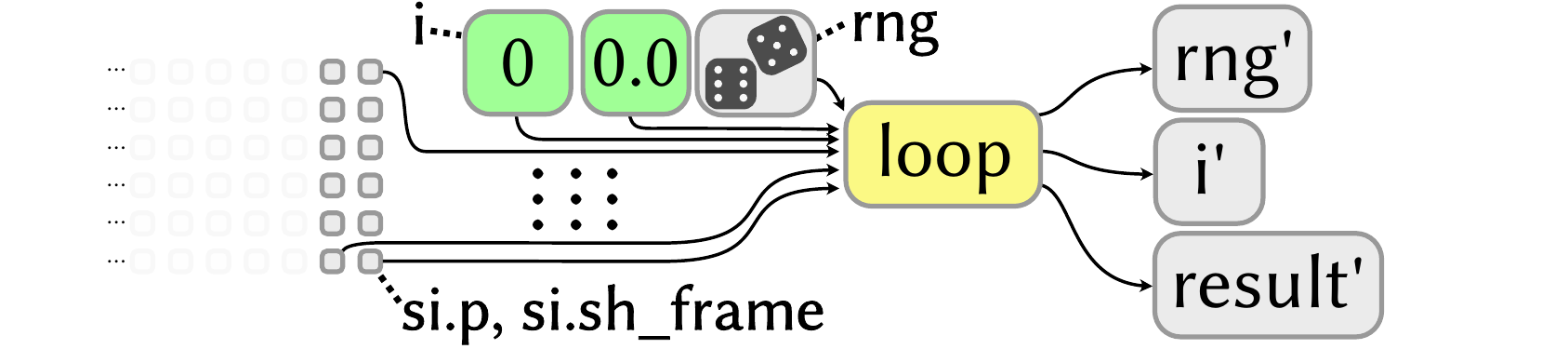}
\end{center}
The specific loop of our example samples the cosine-weighted hemisphere and
traces shadow rays.
\begin{minted}[fontsize=\footnotesize]{python}
while loop(si.is_valid() & (i < 128)):
    # Sample from cosine-weighted hemisphere
    sin_phi, cos_phi = dr.sincos(rng.next_float() * dr.two_pi)
    sin_theta_2 = rng.next_float()
    wo_local = Vector3f(cos_phi * dr.sqrt(sin_theta_2),
                        sin_phi * dr.sqrt(sin_theta_2),
                        dr.sqrt(1 - sin_theta_2))
    # Rotate sample into shading frame and spawn ray with length 1
    ray_2 = si.spawn_ray(si.sh_frame.to_world(wo_local))
    ray_2.maxt = 1
    # Count unoccluded rays and increase the iteration count
    result[~scene.ray_test(ray_2)] += 1.0
    i += 1
\end{minted}
\vspace{.8mm}
\noindent Finally, we reshape and store the result as an image (right). It is
at this point that the actual computation takes place within a megakernel
containing all steps except for \texttt{x} that was previously
evaluated.\\[-1.5mm]
\begin{minipage}[t]{.65\columnwidth}
\vspace{0pt}
\begin{minted}[xleftmargin=0pt,fontsize=\footnotesize]{python}
from mitsuba import Bitmap
# Arbitrary-rank float32 tensor
from drjit.cuda import TensorXf
# Reshape/scale
tensor = TensorXf(result / 128,
                  shape=(1024, 1024))
Bitmap(tensor).write('out.exr')
\end{minted}
\end{minipage}%
    \hspace{2mm}
\begin{minipage}[t]{.29\columnwidth}
\vspace{0pt}
\setlength{\fboxsep}{0pt}%
\setlength{\fboxrule}{1pt}%
\fbox{\includegraphics[width=\textwidth]{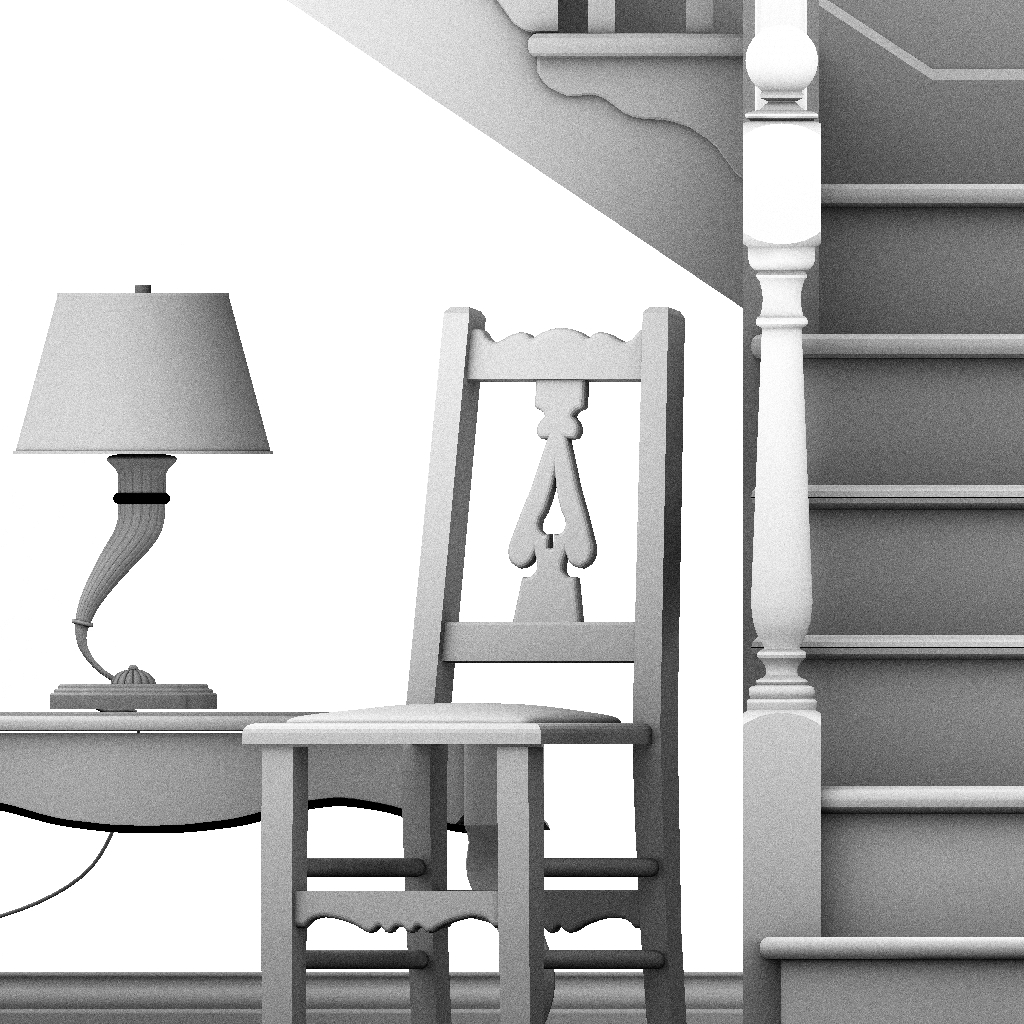}}
\end{minipage}

\subsection{Discussion}
Following this example, we are in a better position to review high-level ideas
of \drjit and contrast it with other approaches.
\vspace{-.3mm}
\begin{enumerate}[leftmargin=6.3mm,label=\arabic*.]
    \item \drjit is principally a framework for tracing large amounts of
        embarrassingly parallel computation.
\end{enumerate}
\vspace{-.3mm}
Tracing constitutes \drjit's only mode of operation and takes place all the
time. Tracing must be highly efficient, since an individual differentiation
step can create millions of temporary JIT variables.
\vspace{-.3mm}
\begin{enumerate}[leftmargin=6.3mm,label=\arabic*.]
    \setcounter{enumi}{1}
    \item  \drjit captures the rendering process
without intermediate evaluation, returning the image in the form of a trace
        describing the computation needed to produce it.
\end{enumerate}
\vspace{-.3mm}
This ultimately enables megakernel compilation and is somewhat specific to
stateless Monte Carlo methods. We did not study methods that build data
structures like photon maps \mbox{or irradiance caches.}
\vspace{-.3mm}
\begin{enumerate}[leftmargin=6.3mm,label=\arabic*.]
    \setcounter{enumi}{2}
    \item Tracing preserves control flow like loops and polymorphism.
\end{enumerate}
\vspace{-.3mm}
Backend compilation time in OptiX and LLVM tends to grow super-linearly with kernel size.
Preserving control flow avoids the creation of an immensely
large unrolled program that breaks the backend.
\vspace{-.3mm}
\begin{enumerate}[leftmargin=6.3mm,label=\arabic*.]
    \setcounter{enumi}{3}
    \item Dynamic compilation enables aggressive specialization
        and simplification of the program representation.
\end{enumerate}
\vspace{-.3mm}
We will shortly see how a complete trace enables scene-specific specialization
and interprocedural optimization.

\vspace{-1mm}
\paragraph{Array programming.}
\drjit shares similarities with array programming frameworks like
\emph{PyTorch}, \emph{TensorFlow}, and \emph{JAX} that are commonly used for
differentiable programming. Like \drjit, some of them can trace loops and
indirect calls and compile them into fused kernels. We implemented a small
renderer using JAX~\cite{Jax} and XLA~\cite{XLA} to run comparisons but were
unable to obtain useful data, as \mbox{compilation timed out on nontrivial
examples.}

Details of how loops and polymorphic operations are traced can greatly impact
compilation and runtime performance, and we further investigated those two
steps in microbenchmarks shown in Appendix~\ref{sec:appendix}. Our take-away
message from these experiments was that the large amount of arithmetic in PBDR
workloads is unusual in the array programming setting designed around neural
computations built from comparatively few arithmetically intensive steps. Rich
ML-centric IR and tensorial optimizations can be highly effective in ML
workloads, but they are costly \mbox{when used to render images.}

\begin{figure*}[t]
    \centering
    \includegraphics[width=\textwidth]{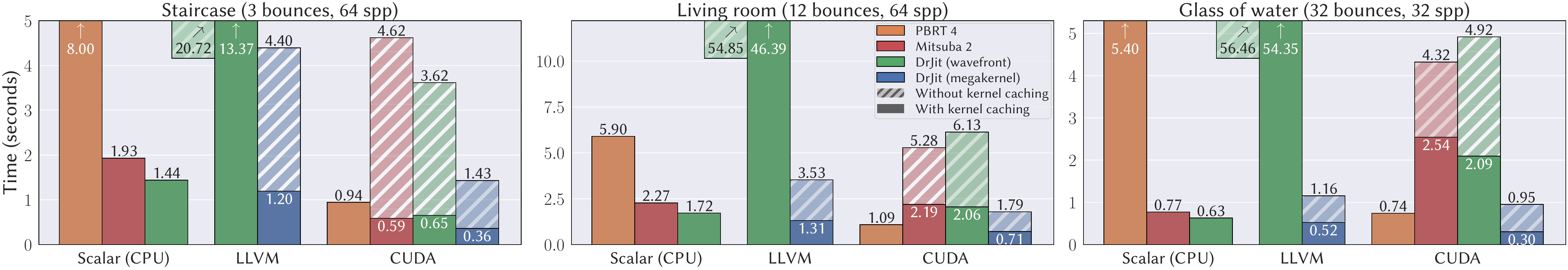}
    \vspace{-6.5mm}
    \caption{%
        \label{fig:benchmark-primal}%
        We compare the performance of the combination of \mitsuba and \drjit with
        and Mitsuba 2~\cite{NimierDavid2019} and
        PBRT 4~\cite{PBRTv4},
        rendering three scenes of varied complexity that are shown in
        Figure~\ref{fig:benchmark-scenes}. Results for both megakernel and wavefront-style
        evaluation are provided for \drjit, where hatched bars indicate the
        one-time backend compilation cost (the actual kernel runtime
        is listed just below). PBRT is statically compiled and
        does not have this overhead. The performance figures demonstrate that
        OptiX and LLVM megakernels produced by \drjit achieve state-of-the-art
        performance.
    }
\end{figure*}

\vspace{-2mm}
\subsection{Optimizations}
\label{sec:optimizations}
\drjit performs optimizations to improve code generation. This may appear
counter-intuitive: LLVM and OptiX are themselves sophisticated optimizing
compilers, hence standard optimization passes would be subsumed by the backend
compilation step. We focus on an important exception: \emph{dynamic dispatch}
presents an opaque boundary that breaks interprocedural optimizations in
LLVM/OptiX.

Recall the large intersection data structure in our running example, of which
only a few fields were used. Since we are already JIT-compiling, why not
generate specialized code that omits unreferenced fields along with the
computation \mbox{needed to produce them?}

This train of thought leads to a set of global optimizations illustrated in
Figure~\ref{fig:vcall-opt}: first, we propagate constant literals into the
call, where it may trigger simplifications (this is especially effective in
differentiated programs that tend to propagate many zero-valued
derivatives). Removing unreferenced outputs removes unreferenced inputs, which
may further cascade into the surrounding program. We also devirtualize
computation that is identical in all sub-traces.

These optimizations are easily performed while tracing: constant propagation
happens automatically, dead code elimination follows from variable reference
counting, and deduplication builds on value numbering: if all
sub-traces output a variable with the same ID, it can be moved out of the call.
Finally, identical traces can be collapsed into a single subroutine. This
optimization is important for rendering, where scenes often contain thousands
of objects of the same type.

However, type equivalence does not imply that two instances will always perform
the same sequence of operations. Consider an \emph{ubershader} material like
Disney's Principled BSDF~\cite{burley2012physically,PBShading} that could
expand into opaque or thin versions with different subsets of the underlying
$\sim$11 parameters active in each instance. Textured parameters could be
procedural, driven by mesh attributes, 2D bitmaps or 3D volumes with various
wrapping or filtering modes. In a spectral renderer, this could furthermore
involve custom spectral profiles or upsampling steps. A subset of these
parameters may require derivative tracking, while others are static. We make no
assumptions and trace every reachable instance, which generally produces many
identical functions that can then be safely merged.

\vspace{-2mm}
\subsection{Additional features}

\paragraph{Vectorization.} The LLVM backend of \drjit generates vectorized IR
targeting the SIMD instruction set of the host. This means that operations like
loops, function calls, and ray tracing process 4-16 elements at a time.
Function calls may require several iterations to handle all elements in the
presence of divergence. \drjit automatically performs such transformations,
while adding masking similar to the \emph{ISPC} compiler~\cite{pharr2012ispc}.
Function deduplication discussed in Section~\ref{sec:optimizations}
is important for vectorized execution on both CPUs and GPUs
because it can significantly reduce divergence.

\vspace{-2mm}
\paragraph{Closure generation.}
\label{sec:closure-gen}
Instance-specific attributes like the roughness parameter of a BRDF model
require special handling during tracing. \drjit automatically detects such
implicit dependences and adds them to a \mbox{\emph{function closure}}.
Appendix~\ref{sec:closures} provides details on this.

\vspace{-2mm}
\paragraph{Loop optimizations.}
Besides polymorphism-related optimizations, \drjit also optimizes loop
constructs by removing invariant or unreferenced state variables. LLVM and
OptiX can in principle perform similar optimizations, but polymorphism within
loops can sometimes impede them. We leverage \drjit's global view across loops
and indirections to remove redundancies.

\vspace{-1mm}
\paragraph{Wavefront-style evaluation.}
\label{sec:wavefronts}
Several flags control the tracing behavior of \drjit: for example, it can
either trace loops or execute them using a sequence of wavefront-style kernels.
Similarly, it can trace polymorphic function calls or group the wavefront by
instance and launch a maximally coherent kernel per group. Besides full
megakernel- or full wavefront-style evaluation resembling Enoki~\cite{Enoki},
this provides an intermediate configuration with wavefront loops and traced
calls that can be surprisingly effective.

\vspace{-1mm}
\paragraph{Side effects.} Several of \drjit's builtin operations
like scatter operations and atomic scatter-reductions mutate existing arrays.
Read access or arithmetic involving variables marked as being affected by
queued side effects trigger a kernel launch to materialize them.
\begin{figure*}[t]
    \centering
    \includegraphics[width=\textwidth]{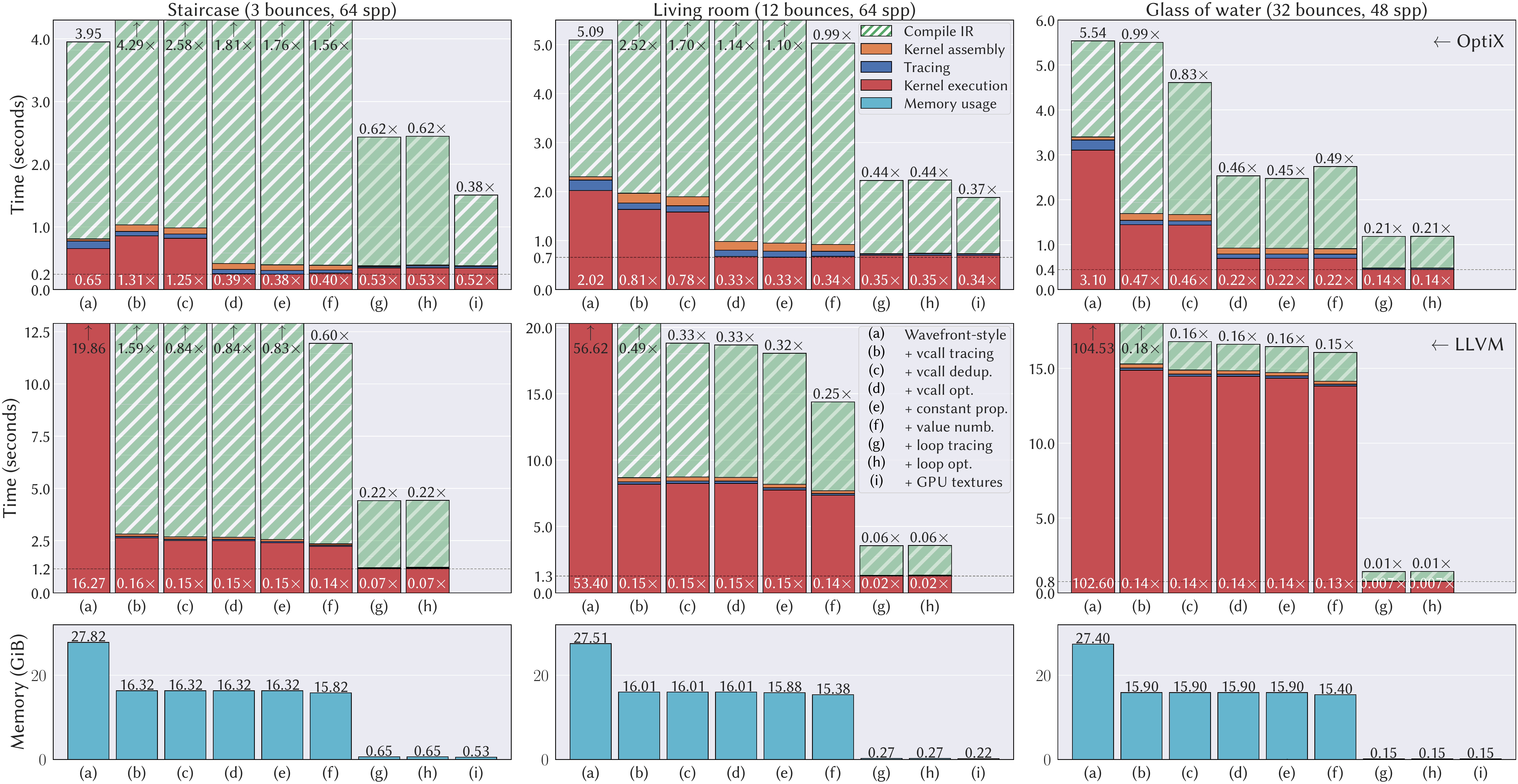}
    \vspace{-6.5mm}
    \caption{
        \label{fig:benchmark-primal-features}%
        We investigate the effect of different
        optimizations in \drjit using the same set of scenes (columns). The rows show OptiX and LLVM runtime
        costs followed by peak memory usage that is identical in both
        backends. Stacked bars indicate the time spent on backend IR
        compilation (hatched), kernel assembly (orange) and tracing
        (blue) within \drjit, as well as kernel execution (red).
        The leftmost bar \textbf{(a)} is a wavefront baseline that ``unrolls''
        loops and polymorphism into separate kernels that
        communicate through device memory.
        Going from left to right, we then successively enable
        \textbf{(b)} compilation of polymorphism into subroutines,
        \textbf{(c)} deduplication of subroutines containing identical code,
        \textbf{(d)} global polymorphism-aware optimizations (Section~\ref{sec:optimizations}),
        \textbf{(e)} constant propagation,
        \textbf{(f)} local value numbering,
        \textbf{(g)} tracing loops instead of unrolling them,
        \textbf{(h)} loop state optimizations, and
        \textbf{(i)} hardware accelerated texture lookups (GPU only).
    }
\end{figure*}

\subsection{Results}
\label{sec:results-1}
We now turn to a first set of results that examine \emph{primal} rendering
performance, comparing \drjit to two open source renderers with wavefront-style
evaluation: Mitsuba 2~\cite{NimierDavid2019} using the
Enoki~library~\cite{Enoki}, and PBRT 4~\cite{PBRTv4}.

\begin{figure}[t]
    \centering
    \includegraphics[width=\columnwidth]{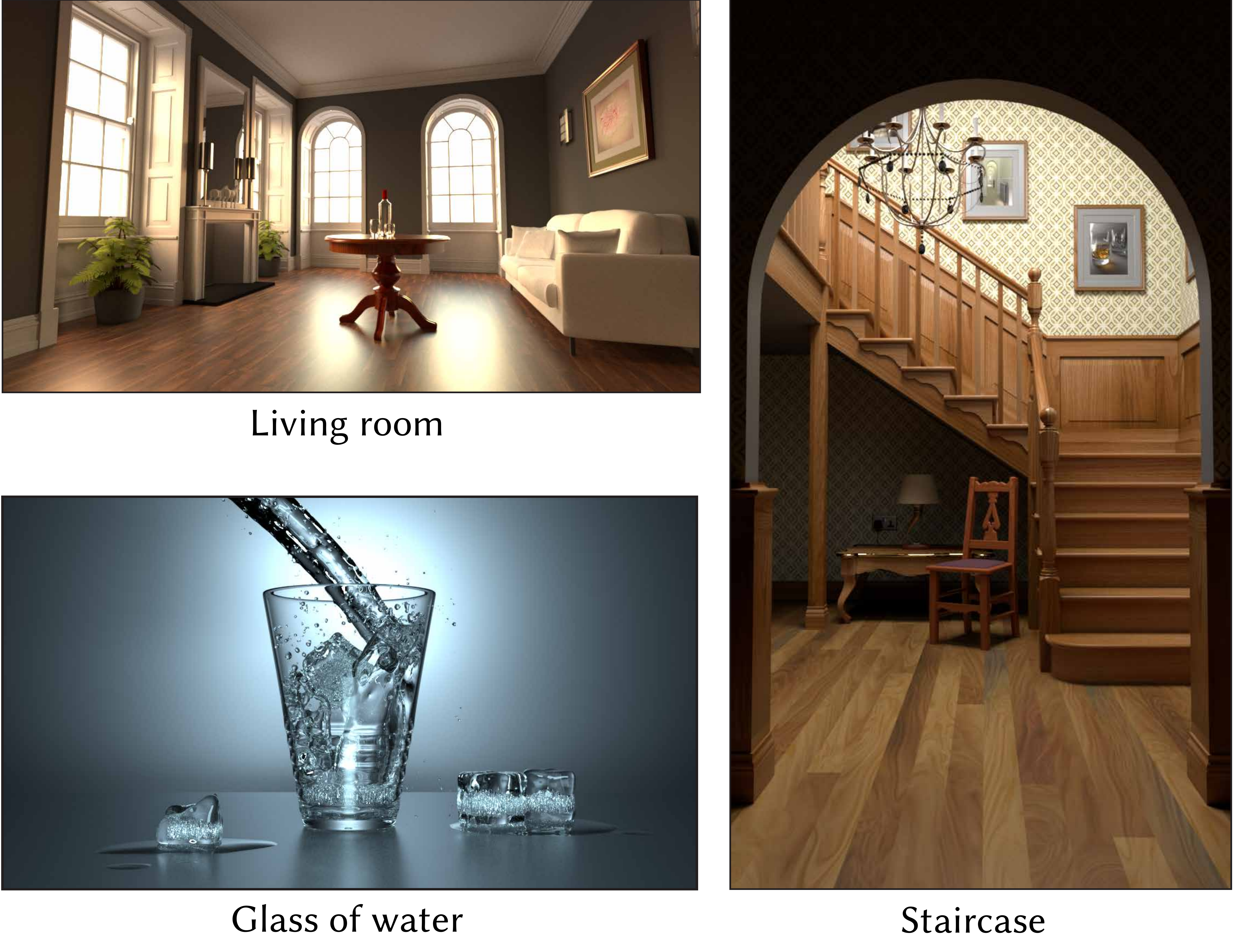}
    \vspace{-7mm}
    \caption{%
        \label{fig:benchmark-scenes}%
        Scenes from Bitterli's~\shortcite{bitterli16resources} rendering
        resources used in benchmarks.
    }
\end{figure}

The three benchmark scenes are shown in Figure~\ref{fig:benchmark-scenes}:
\emph{Staircase} contains 749 shapes and 24 BSDFs and simulates low-order
scattering, while \emph{Living room} uses longer paths with 12 interactions.
\emph{Glass of water} simulates up to 32 dielectric refractions and
reflections. Good performance in these benchmarks requires efficient handling
of loops for interreflection and dynamic dispatch to \mbox{many scene objects.}

Experiments throughout this article were performed on an AMD Ryzen Threadripper
3990X server (64+64 virtualized cores, 128 GiB RAM) with an NVIDIA RTX A6000
GPU (45.6~GiB RAM). The vendor-specified thermal design power of CPU (280W) and
GPU (300W) are roughly matched to enable a fair comparison between these two
very different processor architectures. On the software side, we use Linux
kernel 5.13, LLVM~10, and NVIDIA driver 510 with a vendor-provided fix for a
performance regression found during development (this fix will be part of
future driver releases). We follow standard benchmarking practices like
disabling mitigations for side-channel attacks, reporting the median of
multiple (5) runs, and pre-allocating 2~MiB \emph{huge pages} used by
\mbox{Embree~\cite{Embree} and \drjit.}

Three groups of columns in Figure~\ref{fig:benchmark-primal} contrast
statically compiled one-ray-at-a-time rendering (\emph{scalar} C++ code) with
execution via LLVM, and OptiX. \drjit results are presented in both wavefront
and megakernel mode. Our system compiles a specialized megakernel (or series of
smaller kernels) for each scene, which incurs a one-time overhead represented
by hatched bars; the solid portion below indicates performance once the system
is warmed up. We applied minor optimization to $\mitsuba$ while incorporating
$\drjit$, such as marking primary rays as coherent in Embree. These also
benefit the renderer's statically compiled scalar mode,
which we represent using the green \drjit bar in the ``Scalar (CPU)''
column.

\drjit performance compares favorably, achieving geometric mean speedups of
$3.70\,\times$ (vs.\ Mitsuba 2) and $2.14\,\times$ (vs.\ PBRT 4) on the GPU. On
the CPU, vectorized LLVM megakernels produced by our system outperform scalar
rendering in all cases.

Compilation into wavefronts produces programs that must repeatedly read and
write vast amounts of data, imposing a burden on the processor's memory
subsystem. Difference in memory bandwidth on the CPU ($\sim$95 GiB/s) and GPU ($\sim$715
GiB/s) as well as latency hiding-capabilities in the latter lead to marked differences
between architectures in this case. Wavefront-style execution reduces GPU
performance by a factor of $2.0-7.7\times$, while the same experiment on the
CPU can increase cost by a factor of more than $100\times$.

Figure~\ref{fig:benchmark-primal-features} investigates the effect of
individual optimizations starting from a wavefront-style baseline
resembling the operation of the Enoki library
(column \textbf{a}). Memory usage is significant in
this configuration (bottom row). While this could be reduced by
launching many smaller wavefronts, it would not address the fundamental issue
that a large amount of data will need to be read and written.

A high level observation is that tracing and kernel assembly (orange and blue
bar regions) only constitute a small portion of the total computation time. The
precise amount tends to be lower when compiling megakernels and higher for
wavefronts requiring multiple rounds of tracing and
assembly. Backend compilation time to transform IR into machine code can be
significant (longer than the kernel execution itself), which can make the
system unsuitable for certain use cases and emphasizes \mbox{the importance of
caching mechanisms.}

Mapping polymorphic function calls to subroutines in \textbf{(b)} drastically
reduces runtime on the CPU while increasing compilation time on both backends;
subroutine deduplication in \textbf{(c)} reverts most of the growth in compilation
time and also gives a speed boost thanks to reduced divergence under vectorized
execution.

Removing unreferenced computation, arguments, and return values from
polymorphic method calls in \textbf{(d)} has a surprisingly large effect in the
OptiX backend, where it improves performance by an average factor of
$2.5\times$, while having essentially no impact on the CPU. Modern superscalar
CPUs can absorb a certain amount of redundant computation by issuing multiple
instructions per clock cycle. We speculate that these architectural features
along with the lower cost of data exchange through the stack could be
responsible for the modest benefit of this optimization (though it will prove
to be more effective when applied to differential kernels in
Section~\ref{sec:system-ad}).

Loop tracing in \textbf{(g)} reduces memory traffic to a minimum, which greatly
benefits CPU execution, while having an inconclusive effect on the GPU. We
speculate that loops may exacerbate the conversion into an \mbox{OptiX} state
machine, whose nodes must exchange continuation state~\cite{OptiX}. Loop
optimizations in step \textbf{(h)} remove~$\sim$65\% of state variables
(details in Section~\ref{sec:results-2}), but this appears to be mostly
subsumed by the backend's optimization passes.

Performing GPU texture lookups using hardware texture mapping units in
\textbf{(i)} yields small but measurable runtime improvements on the order of
1\% when compared to software-based interpolation. Compilation time improves by
up to 30\% when the scene contains many textured materials (e.g.,
\textit{Staircase}), since IR operations related to wrapping and interpolation
can be removed.

\vspace{-1.5mm}
\subsection{Limitations}
The following are notable limitations of the tracing approach.

\vspace{-1mm}

\paragraph{The need to evaluate.} Suppose we wanted to run the ambient
occlusion integrator multiple times to produce uncorrelated renderings. A
na\"ive attempt to do so by reusing the pseudorandom number \texttt{rng} could
lead to a problem where caching breaks down, requiring repeated backend
compilation of kernels that become bigger and slower over time. This happens
because we never asked the system to materialize the RNG state into data; all
RNG updates from prior rendering steps are still part of the trace and spill into the current kernel. Calling \mbox{\texttt{dr.eval(result,\!\! rng)}} solves this
problem, though the need for this step can be non-intuitive to new users.

\paragraph{Recursion.}
\drjit can trace dynamic dispatch to functions that themselves perform dynamic
dispatch, which is, e.g., needed to handle object transformations involving
instanced geometry. Cycles including self-recursive calls are not
permitted, as the system would then trace the same
code repeatedly without knowing when to stop.

\paragraph{Debugging.}
We provide two ways to investigate the behavior of a program: the user
can set \drjit to wavefront-style evaluation (Section~\ref{sec:wavefronts}),
set breakpoints, step through the program, and investigate variable contents.
Megakernel compilation is less flexible: the only available option in this case
is to print from the device via \texttt{dr.printf\_async()}, which will produce
output when the kernel runs later. More work will be needed to develop better
debugging primitives for this unusual way of performing computation.

\section{Differentiation}
\label{sec:system-ad}
Following this presentation of \drjit's computational foundation, we now
turn to differentiation starting with a review of PBDR techniques and ways
in which they constrain the system.

\begin{figure*}[t]
    \centering
    \includegraphics[width=\textwidth]{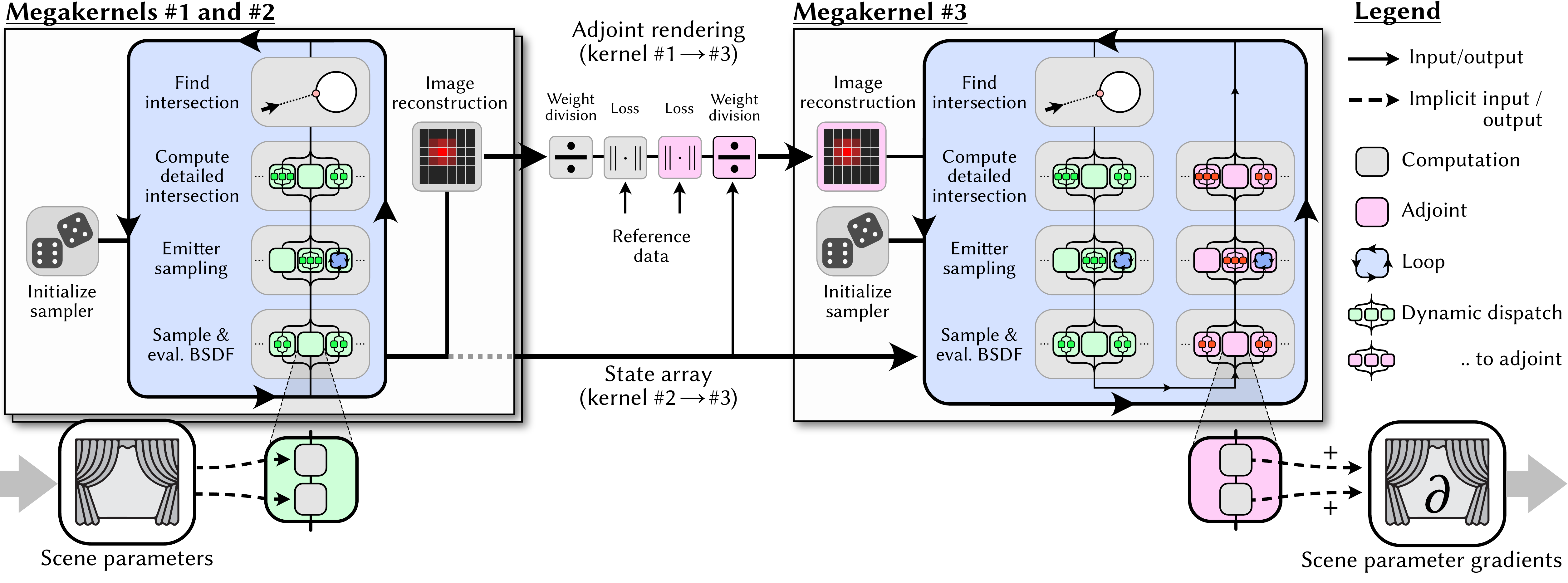}
    \vspace{-4.5mm}
    \caption{
        The anatomy of a recent physically based differentiable rendering
        method. This diagram illustrates a partition of the major components of
        \emph{Path Replay Backpropagation}~(PRB)~\cite{Vicini2021} into a set
        of three self-contained megakernels that each solve a Monte Carlo
        integration problem. Megakernel~\#1 is a standard path tracer that
        dynamically dispatches function calls to scene objects (shapes,
        emitters, materials, etc.) with an implicit dependence on scene
        parameters. The path tracer uses an image reconstruction filter to
        scatter weighted samples values into an output buffer with subsequent
        weight division~\cite{PBRTv4}, which produces a \emph{primal} rendering
        that is passed into a loss function to quantify the quality of the
        current iterate. Meanwhile, a second path tracing megakernel
        performs the same computation once more with a different pseudorandom
        seed, providing a per-sample state array to a final \emph{adjoint}
        megakernel \#3. This kernel contains adjoint versions of all steps and
        is responsible for accumulating scene parameter gradients.
    }
    \label{fig:prb-structure}
\end{figure*}

\subsection{Physically Based Differentiable Rendering}
\label{sec:pbdr-review}
A typical rendering algorithm like path tracing~\cite{KajiyaRenderingEquation}
propagates \emph{radiance} from light sources to the sensor. To do so, it loops
over consecutive path vertices starting from the sensor. Many iterations might
be necessary when the scene contains scattering media.

Such loops are unfortunately a nuisance during differentiation, as each
iteration generates intermediate state that must be reconstituted to compute
reverse-mode derivatives~(Section~\ref{sec:source-trafo}). In the case of a
path tracer, millions of Monte Carlo samples will run the loop in parallel for
an unpredictable number of iterations, hence large quantities of memory would
need to be provisioned conservatively to store the resulting intermediate
state. In general, we find that the automatic derivative of a loop is almost
never \mbox{satisfactory}; the user should instead contribute domain-specific
knowledge to implement an equivalent and more efficient \emph{custom adjoint}.

\emph{Radiative Backpropagation}~(RB)~\cite{NimierDavid2020} is such a custom
adjoint. Using physical reciprocity, it replaces the derivative of the
simulation with an equivalent simulation of \emph{differential} radiation
(``\emph{adjoint radiance}''), removing the need to store loop state.
Differentiating the image loss initially yields adjoint radiance in image
space, where it describes how pixels in the rendered image should change to
reduce the loss. Like a video projector, the camera then emits this signed
radiation into the scene, where it scatters just like normal light. In contrast
to primal rendering that queries the scene representation using \emph{reads},
RB is a \emph{write}-heavy method: whenever adjoint radiance encounters a
surface with differentiable parameters, the method accumulates a contribution
into the local parameter gradient, for example to optimize $\alpha$ below.
\begin{center}
    \vspace{1mm}
    \includegraphics[width=1\columnwidth]{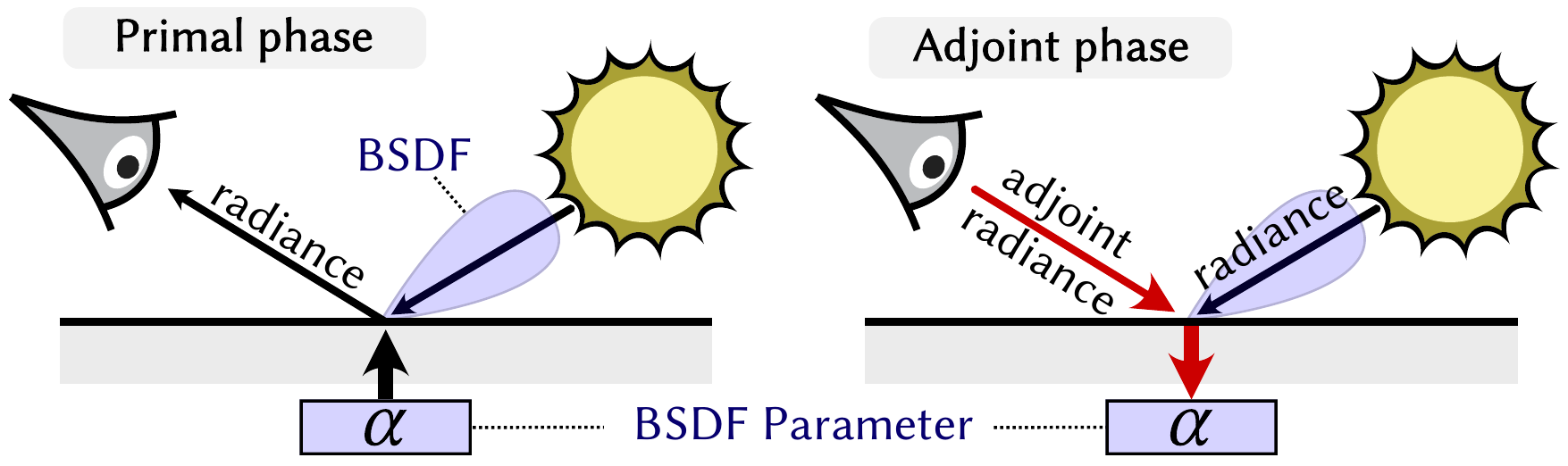}
\end{center}
The derivative with respect to $\alpha$ is also proportional to the amount of
incident radiance (a surface in darkness generates no gradients). This
illustrates the main issue with RB: it requires an intertwined simulation of
both adjoint radiance \emph{and} primal radiance, which leads to a costly
algorithm with quadratic runtime complexity.

\emph{Path Replay Backpropagation}~(PRB)~\cite{Vicini2021} fixes this problem
using a two-pass approach. Its first pass generates a set of Monte Carlo
samples and stores data consumed by a subsequent pass. This second pass
regenerates the same set of samples, exploiting the precomputed data and
arithmetic invertibility to recover the incident radiance at every vertex. With
this information, the desired gradients can be accumulated. A complete PRB
optimization step actually computes one further
\emph{decorrelated}~\cite{gkioulekas2016evaluation} primal rendering that is
provided to the image loss function, so there are \emph{three} integration
steps in total. Figure~\ref{fig:prb-structure} illustrates the high-level
architecture and a suggested decomposition into megakernels.

\begin{figure}[b!]
        \vspace{.5mm}
    \centering
    \includegraphics[width=\columnwidth]{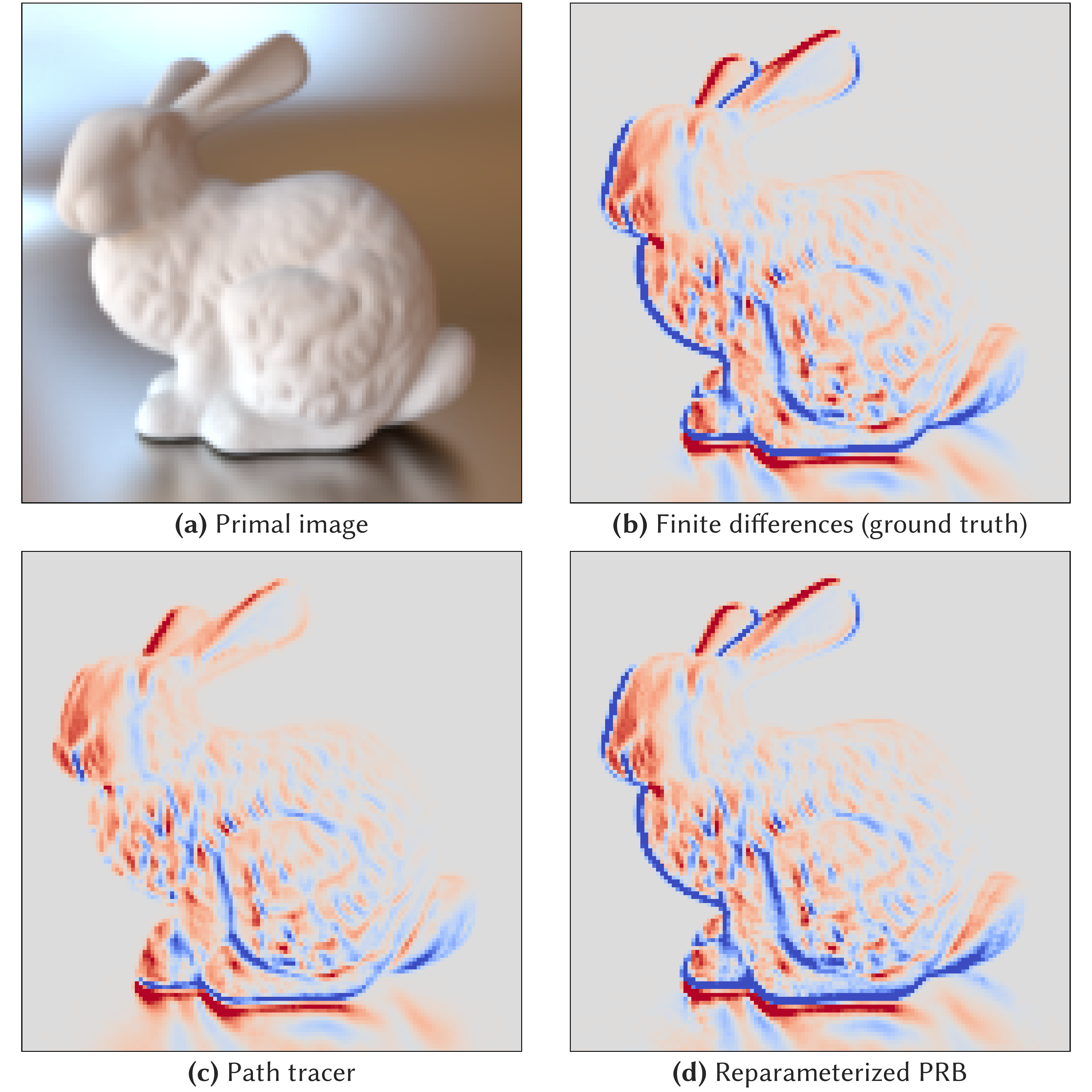}
    \vspace{-6mm}
    \caption{%
        \label{fig:bunny}%
        Parameters that affect the visibility in a scene produce biased
        derivatives unless extra precautions are taken. This figure shows the
        forward derivative of a Stanford bunny on a
        metallic floor with respect to translation. \mbox{\textbf{(a)} Primal rendering.} \textbf{(b)} Reference.
        \textbf{(c)}~Na\"ive derivative of a path tracer. Bias manifests in the
        form of incorrect silhouette gradients and reflections in the
        floor. \textbf{(d)} Reparameterizing integrals removes the discrepancy.
    }
\end{figure}

There are other important problems to consider besides efficiency. Rendering
algorithms evaluate integrals containing visibility-related discontinuities.
When differentiated parameters affect visibility (and thereby the position of
the discontinuities), the resulting derivatives tend to be severely biased,
which breaks shape reconstruction unless special precautions are taken.
Figure~\ref{fig:bunny} visualizes bias in a simple forward mode example.
Techniques to remove bias include \emph{edge
sampling}~\cite{li2018differentiable}
\emph{reparameterizations}~\cite{Loubet2019,Bangaru2020}, and \emph{path space
sampling}~\cite{zhang2020path}.

Finally, differentiation changes the integrals computed by the simulation.
Sampling strategies designed for a primal integrand may perform poorly when
applied to its derivative, in which case derivative-aware strategies may be
necessary~\cite{Zeltner2021MonteCarlo}.
\begin{figure*}[t]
    \centering
    \includegraphics[width=\textwidth]{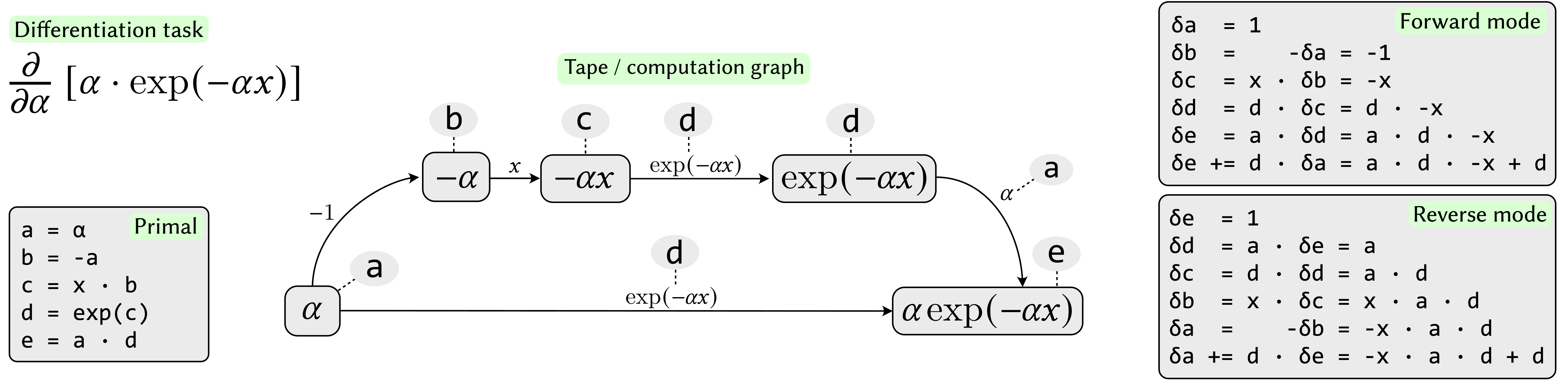}
    \vspace{-6.5mm}
    \caption{%
        \label{fig:ad-tape}%
        An example use of tape-based forward/reverse-mode AD to differentiate
        the exponential density $\alpha\exp(-\alpha x)$ with respect to
        $\alpha$. The variable $x$ does not carry derivatives (it is
        \emph{detached}) and could, e.g., represent a Monte Carlo sample
        position. Primal evaluation of this expression (bottom left) generates
        a sequence of temporaries (\texttt{a} \texttt{b}, \texttt{c},
        \texttt{d}, and \texttt{e}) and replicates the dependency structure of
        their computation in a tape/computation graph (middle). Each edge in
        this graph carries a weight indicating the sensitivity of the
        target node with respect to perturbations of the source node. Often,
        these edge weights are simply variables of the primal program, which is
        indicated by ellipses. AD associates a derivative variable with each
        temporary and program variable (e.g., \texttt{a} and
        \texttt{$\delta$a}). Forward- and reverse modes (right) traverse the tape as
        indicated by their name, propagating derivatives scaled according to
        the edge weights. The result is equivalent, though the efficiency of
        the two modes can drastically differ when the computation has many
        inputs or outputs.
    }
\end{figure*}

Our goal in the remainder of this paper is to establish a solid foundation for
these and future PBDR methods. We must, however, limit the scope somewhat and
therefore mainly focus on the example of a reparameterized path replay
backpropagation integrator.

\subsection{Objectives}
A central tenet of Sections~\ref{sec:intro} and~\ref{sec:system-jit} was that
the system should never partition Monte Carlo integration across multiple
kernels to avoid costly inter-kernel communication. Tracing also needed to
preserve loops and polymorphic calls to avoid creating immensely large unrolled
kernels that would be challenging to compile.

The same objectives also apply to the generation of \emph{differential
megakernels}. \drjit must be able to trace an algorithm like PRB and compile
its Monte Carlo integration steps into self-contained kernels with preserved
control flow. This introduces several new challenges related to
differentiation:

\begin{itemize}[leftmargin=6.3mm]
    \item PRB performs a differentiable computation at every scattering
        interaction and then invokes AD to back- or forward-propagate
        associated derivatives. \drjit must convert these high-level AD operations
        into code that can be included in a megakernel.\\[-2.6mm]

    \item The differentiable computation contains many polymorphic
        function calls. Their derivative should also be a polymorphic call
        targeting generated forward or reverse-mode derivative versions of the
        primal function implementations.\\[-2.6mm]

    \item Derivatives should flow through typical preprocessing
        steps like the computation of smooth normals or MIP maps.
        However, data dependencies prevent the evaluation of these derivatives
        within the same megakernel. \drjit must partition the AD task into
        multiple phases \mbox{with efficient information exchange.}\\[-2.6mm]

    \item \drjit must support derivative-related transformations that transcend raw
        differentiation. For example, it should be possible to introduce
        complex transformations like reparameterizations.
\end{itemize}

\drjit once more addresses these requirements using an approach based on
tracing. This tracing of differentiable computation occurs at a higher
architectural level and depends on the JIT compiler as foundation, which means
that the combined system traces at \emph{two levels simultaneously.} This
combination is harmonious: besides fulfilling the stated requirements, it
enables \drjit to dynamically specialize differential algorithms to the scene
and problem statement, while removing redundant steps introduced by the AD
transformation.

A caveat worth noting here is that \drjit does not free the developer from
thinking about subtle details of the differentiation process. It
facilitates---but does not automate---the introduction of the previously
mentioned physical and mathematical ``reversibilities''.

The remainder of this section reviews tracing AD and explains how it can be
adapted to address the stated challenges. Unless noted otherwise, subsequent
use of the word \emph{trace} will now refer to a \emph{(Wengert) tape}, or
\emph{computation graph} that captures operations for subsequent derivative
propagation.

\subsection{Tracing with dynamic compilation}
\label{sec:tracing-dyn}

Section~\ref{sec:related} introduced two high-level flavors of AD:
\emph{tracing} and \emph{source transformation}. \drjit lies somewhere in
between: it ingests computation using tracing, but the combination with dynamic
compilation and function-level differentiation produces output resembling that
of source transformation AD.

Figure~\ref{fig:ad-tape} shows a simple example of tracing-based AD. The tape
associates every primal variable $x$ with a differential version $\delta x$ and
records inter-variable dependencies. Subsequent tape traversal computes the
differentiable variables, which produces a flurry of operations referencing
primal program state (Figure~\ref{fig:ad-tape} right). A subtle but important
detail is that this derivative-related computation is itself \emph{traced} by
the underlying JIT compiler. Said overly dramatically, differentiation destroys
the AD data structures, which leaves a residue of ordinary computation. This
residue is traced by the JIT-compiler one level below so that it can run as
part of a (mega-) kernel at some later point. The actual destruction of the
tape turns out to be immaterial in this process; it is often useful to retain
it when multiple AD traversals are needed. This approach is not a contribution
of \drjit and also underlies other hybrid AD systems, e.g.,
CppADCodeGen~\cite{CppADCodeGen}.

It is interesting to note that this combined system supports
\emph{checkpointing} without having put any intentional effort into realizing
such a feature. Consider an unrolled \mbox{loop with intermediate evaluation:}
\begin{minted}[fontsize=\footnotesize]{python}
for i in range(1000):
    data = f(data) # 'f' represents a potentially complex
    dr.eval(data)  # transformation of 'data'.
dr.backward(data)  # Now, backpropagate through all steps
\end{minted}
\noindent The AD layer will reference intermediate steps and temporaries produced by
the function \texttt{f} to enable subsequent differentiation, and these must be
recomputed. The JIT layer knows how to compute these needed variables, and it
will do so from the last evaluation point that therefore takes the role of a
checkpoint. Checkpointing is not ideal for PBDR workloads due to the size of
even a small number checkpoints, though it does help when differentiating
rendering techniques that have not been adapted for efficient differentiation.

\vspace{-1mm}
\subsection{Custom adjoints} 
Like other AD systems, \drjit supports functions with user-provided forward-,
and reverse-mode derivatives (``custom adjoints''), by extending an interface
named \texttt{dr.CustomOp}. When an AD traversal encounters a \texttt{CustomOp}
on the tape, it will invoke its \texttt{forward()} or \texttt{backward()}
callbacks to convert function input derivatives into output derivatives, or
vice versa. In contrast to other AD systems that consider \emph{explicit}
function arguments and return values, \drjit also tracks \emph{implicit} inputs
or outputs. This refers to dependencies on global state that only become
apparent while running the function. Tracking implicit reads and writes ensures
that the global flow of derivatives is correctly represented and will be useful
shortly.

\subsection{Differentiating polymorphism}
With this infrastructure in place, we are ready to discuss
polymorphism. We will consider the following method call,
where \texttt{obj} refers to a \drjit
array of instances that implement the function \texttt{func}.
\begin{minted}[fontsize=\small]{python}
out = obj.func(arg_1, arg_2, ..)
\end{minted}
\noindent The derivative of such a polymorphic method call, is \emph{another} polymorphic method call
to JVP and VJP (Section~\ref{sec:source-trafo}) versions of the original set of
functions.
\drjit generates these dynamically as needed. To detect this need, it wraps use
of polymorphism like the example above into a dynamically generated
\texttt{dr.CustomOp}.
Suppose that the $\texttt{CustomOp.backward()}$ callback is now invoked during an
AD traversal, which indicates that the system needs to propagate
\texttt{dr.grad(out)} to
\texttt{dr.grad(arg\_1)},
\texttt{dr.grad(arg\_2)}, etc.

To accomplish this, the AD layer dynamically defines a VJP per
method, which is simply a placeholder that
calls \texttt{func} a second time, with recursive usage of AD to
propagate and extract derivatives.
\begin{minted}[fontsize=\footnotesize]{python}
def func_vjp(grad_out, arg_1, arg_2, ..):
    dr.enable_grad(arg_1, arg_2, ..)
    result = func(arg_1, arg_2, ..)
    dr.set_grad(result, grad_out)
    dr.backward(result)
    return dr.grad(arg_1), dr.grad(arg_2), ..
\end{minted}
\noindent The AD layer then issues a polymorphic method call to these newly
defined functions. To realize this method call, the JIT layer traces each possible target, which
runs all VJP placeholder functions, materializing the differentiation
into ordinary traced computation.

\paragraph{Optimizing polymorphic derivatives.}
Suppose that one of the return values of \texttt{obj.func()} does not carry
derivatives. \drjit will devirtualize this zero-valued derivative following
Section~\ref{sec:optimizations}, enabling optimizations on the caller's side.
Zero-valued input derivatives are propagated into the call, simplifying call
targets using constant propagation. The effect of these simple steps can be
significant.

\paragraph{Automatic discovery of implicit dependences.}
Section~\ref{sec:closure-gen} and Appendix~\ref{sec:closures} observed that
the functions being differentiated are generally \emph{closures} that
implicitly depend on further variables from a surrounding environment, for
example BRDF or texture parameters. This creates differentiable dependences
that must be tracked by the AD system. Our system monitors variable accesses
during the primal \texttt{CustomOp} evaluation and automatically registers
implicit variable dependences with the previously discussed mechanism.

The reverse mode derivative of a method call with implicit reads will issue
atomic scatter-reductions to update scene parameters gradients. This is not
just an odd corner case; it constitutes the main way in which scene parameter
gradients are generated.

\begin{figure}[t]
    \centering
    \includegraphics[width=\columnwidth]{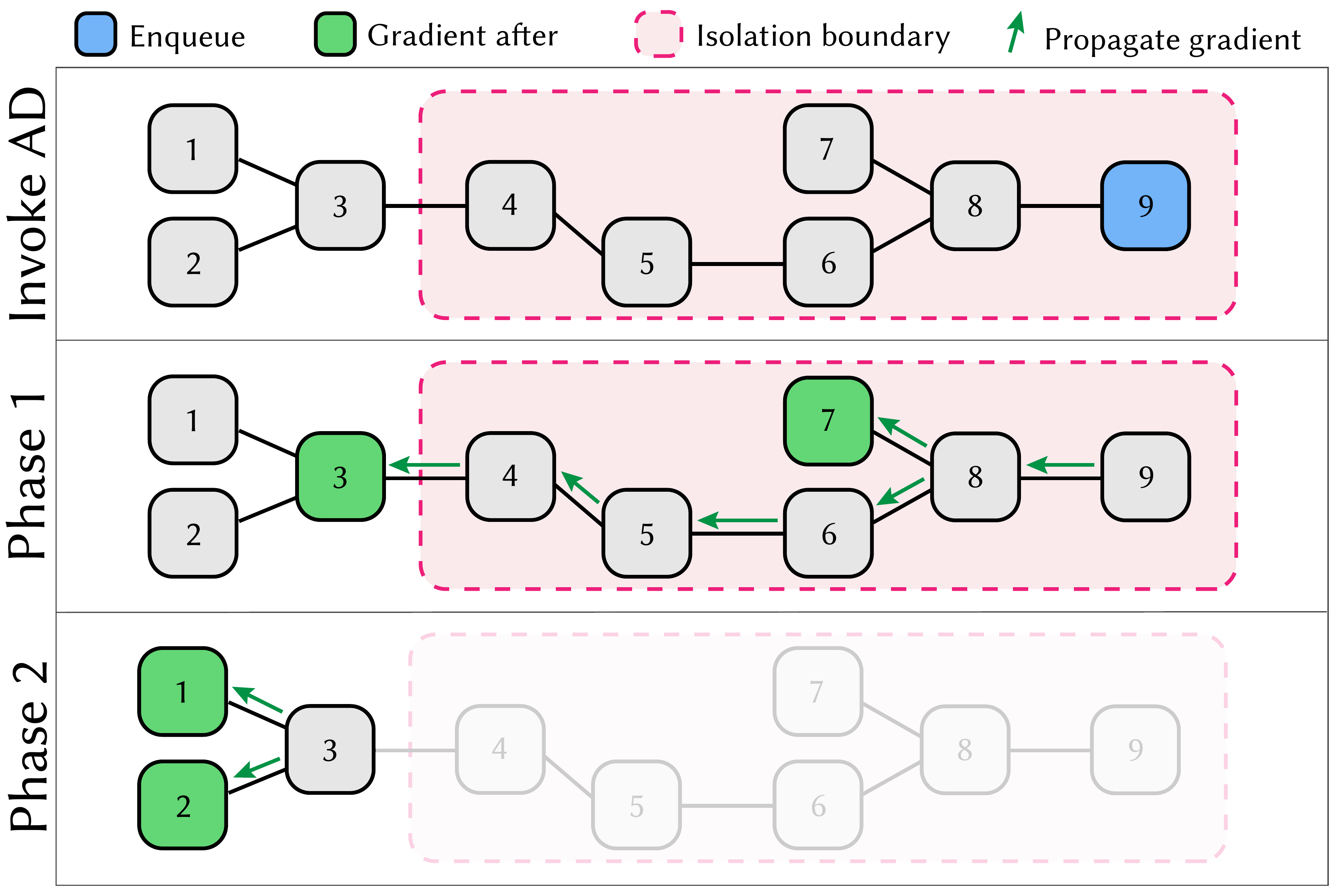}
    \vspace{-6mm}
    \caption{%
        \label{fig:ad-isolate}%
        \drjit provides three different AD \emph{scopes}. The first is an
        \emph{isolation boundary}. Wrapping computation into such a scope
        restrains the process of derivative propagation: only edges within and
        just across the isolation boundary may be traversed, while others are
        postponed until the scope is destructed. \drjit uses this feature to
        ensure correct generation of a differential megakernel that is embedded
        within a larger differential calculation.
    }
\end{figure}
\vspace{-1mm}
\subsection{Isolation boundaries}
PRB performs a differentiable computation at every scattering interaction,
followed by a backpropagation step.
In the example below,
\texttt{Li}, \texttt{Lr} and
\mbox{\texttt{δL} refer to incident, reflected, and adjoint radiance.}
\begin{minted}[fontsize=\footnotesize]{python}
while loop(depth < max_depth):
    # ... Compute surface interaction 'si '...
    Lr = Li * si.bsdf().eval(si, wo) # Reflected radiance from 'wo'
    dr.backward(δL * Lr) # Backpropagate product
\end{minted}
\noindent
This differentiation step
is likely part of a larger computation
including preprocessing steps that must also be differentiated.
Suppose the
that BSDF queries a texture, whose MIP
levels were computed using successive reductions: the
derivative of this process must upsample gradients back to the
finest level, requiring multiple passes with intermediate data
exchange, which cannot be done within the megakernel being compiled. \drjit
provides \emph{isolation boundaries}~(Figure~\ref{fig:ad-isolate}) to
\emph{postpone} such problematic steps.
\begin{minted}[fontsize=\footnotesize]{python}
with dr.isolate_grad():
    # .. temporarily isolate outside world from AD traversals ..
\end{minted}
\noindent Loops and polymorphic calls implicitly create an
isolation boundary.

\subsection{Reparameterizations}
Parameters influencing the position and shape of scene geometry produce bias
when differentiated (Figure~\ref{fig:bunny}) unless the Monte Carlo integrals
within the rendering algorithm are reparameterized. Reparameterizations
counteract parameter-dependent silhouette motion so that
discontinuities are \emph{frozen} when observed within spherical integrals.
We expose
reparameterizations through a \texttt{CustomOp}-based abstraction to isolate
their specifics from the PBDR integrator. A
reparameterized ray intersection then reduces to two lines:
\begin{minted}[fontsize=\footnotesize]{python}
ray.d, det = reparameterize(ray)
si = scene.ray_intersect(ray)
\end{minted}
\noindent
where \texttt{det} is Jacobian determinant of the change of variables.

\begin{figure}[t]
    \centering
    \includegraphics[width=\linewidth]{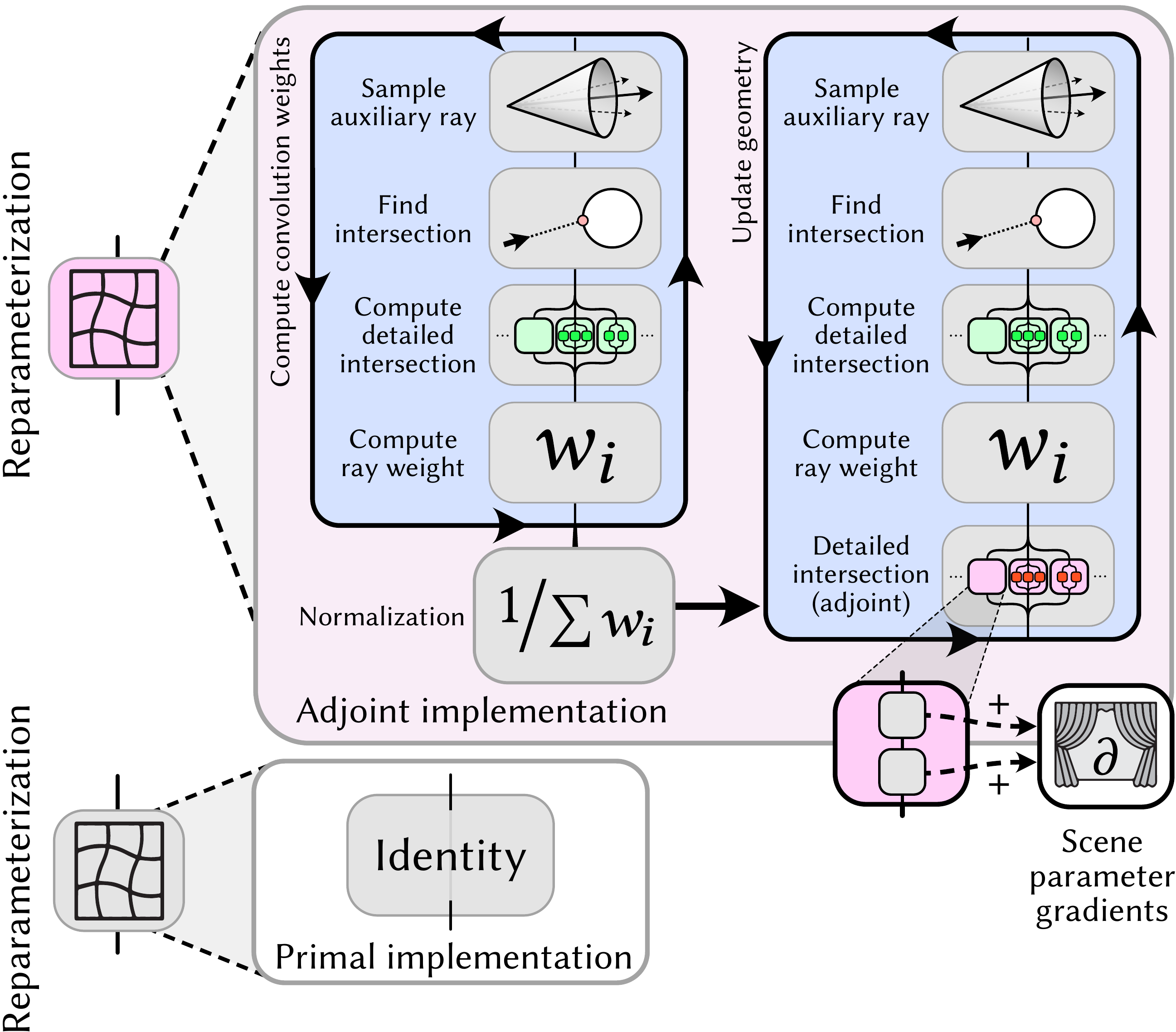}
    \vspace{-5mm}
    \caption{
        To address the visibility-induced bias shown in Figure~\ref{fig:bunny},
        rays must be \emph{reparameterized}~\cite{Loubet2019}. In primal
        computation, reparameterizations have no effect and reduce to the
        identity. When differentiated, they
        trace auxiliary rays to construct a warp
        field~\cite{Bangaru2020} that counteracts silhouette motion. The
        reverse-mode derivative of this warp field scatters scene parameter
        gradients into intersected geometry.
    }
    \label{fig:prb-reparam-structure}
\end{figure}

Reparameterizations are a somewhat strange concept. During primal
rendering, they reduce to the identity: \texttt{reparameterize()}
simply returns the input ray direction along
with a Jacobian determinant of 1.
When differentiated, they expand into intricate
operations
(Figure~\ref{fig:prb-reparam-structure}) that entail tracing auxiliary rays,
weighting them, and using AD to scatter
gradients into intersected shapes.
Usage of AD is therefore recursive: a top-level \texttt{render()}
function launches PRB (a \texttt{CustomOp}) when differentiated, which
differentiates contributions in a traced loop and thereby evaluates the
derivatives of various system operations including parameterizations (a
\texttt{CustomOp}), which launches differentiable ray tracing operations (a
\texttt{CustomOp}). Another level of recursion may take place in the presence
of instancing.
\begin{center}
    \includegraphics[width=.8\columnwidth]{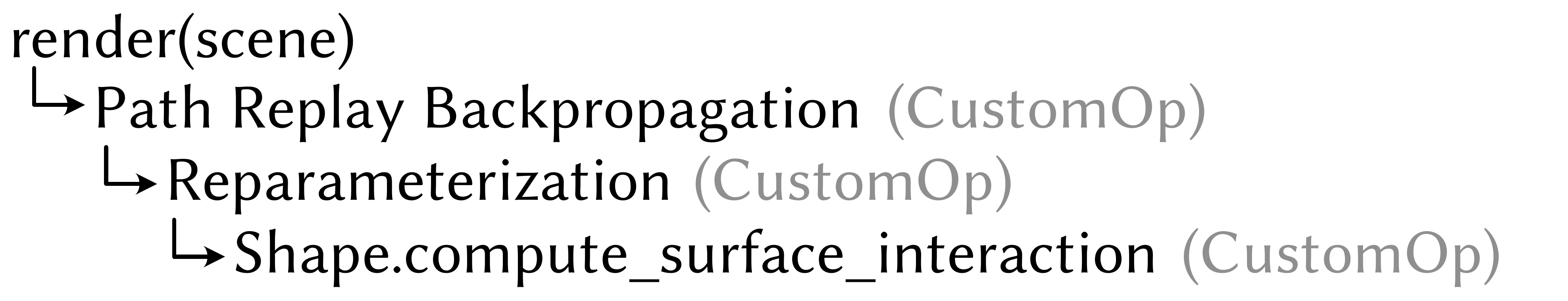}
\end{center}
\subsection{Selective evaluation of partial derivatives}
\label{label:fine-grained}
PRB consists of a loop with nested use of AD to compute light path derivatives
one vertex at a time. Reparameterizations introduce an extra complication by
perturbing the positions of these path vertices.
\begin{center}
    \vspace{-1mm}
    \includegraphics[width=.8\columnwidth]{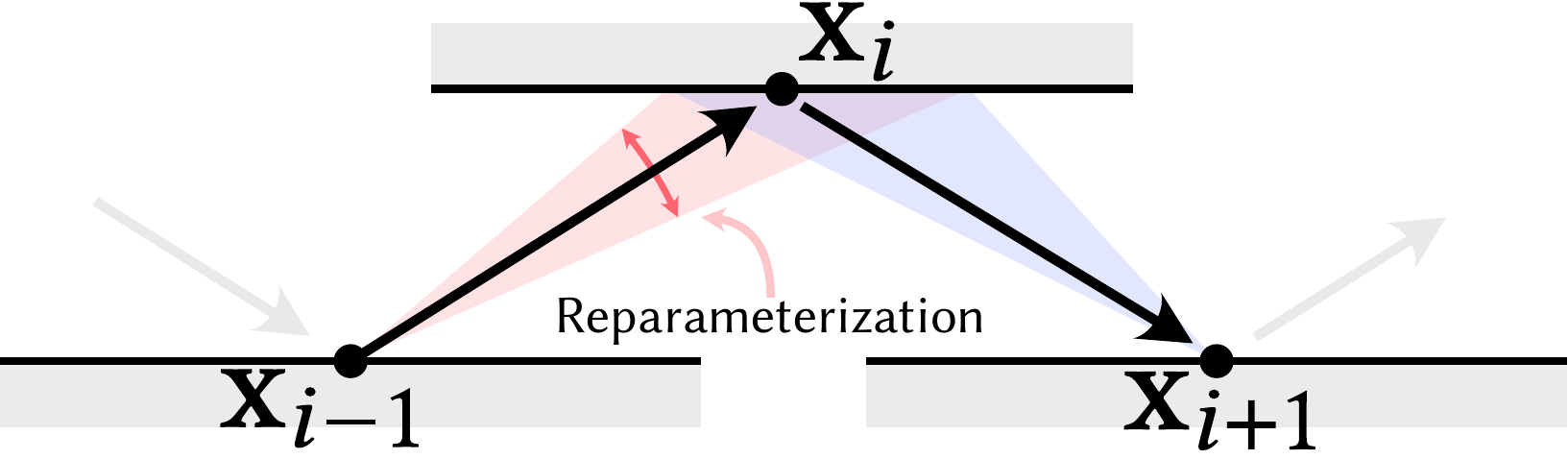}
    \vspace{-1mm}
\end{center}
Moving $\mathbf{x}_{i}$ may change the
BSDF at $\mathbf{x}_{i-1}$ and $\mathbf{x}_{i+1}$.
A reparameterized integrator must account for this dependence
when visiting the reparameterized vertex $\mathbf{x}_{i}$ to ensure that all
partial derivative terms are present.
However, separate differentiation of these two BSDF terms
would also generate scene parameter derivatives, e.g., with respect to the
albedo of $\vx_{i-1}$ and $\vx_{i+1}$. This is undesirable
when those derivatives were already generated elsewhere:
we require a way of only differentiating the dependence on the position of $\vx_i$.

To address these and similar issues, \drjit's AD layer keeps track of a set of
variables $\Omega$ for which derivative tracking is currently enabled. Two
additional \emph{AD scopes} modify this set, by setting it to the empty set
$\emptyset$, its complement $\emptyset^c$, or by adding and subtracting variables from the
current set. A PBDR algorithm using these abstractions
repeatedly enters and leaves scopes to control what derivative terms
should be generated.
\begin{minted}[fontsize=\footnotesize,escapeinside=||,mathescape=true]{python}
with dr.suspend_grad():                   # Ω = $\emptyset$
    with dr.resume_grad():                # Ω = $\emptyset^c$
        ray.d, det = reparameterize(ray)
        si = scene.ray_intersect(ray)
    # .. detached sampling steps ..
    with dr.resume_grad(ray.d):           # Ω = {ray.d}
        # Only account for directional derivatives at 'si_prev'
        L += Li_prev * si_prev.bsdf().eval(si_prev, ray.d)
    # .. other steps ..
    with dr.resume_grad():                # Ω = $\emptyset^c$
        dr.backward(δL * L) # Backpropagate through all terms
\end{minted}

\begin{figure*}[t]
    \centering
    \includegraphics[width=\textwidth]{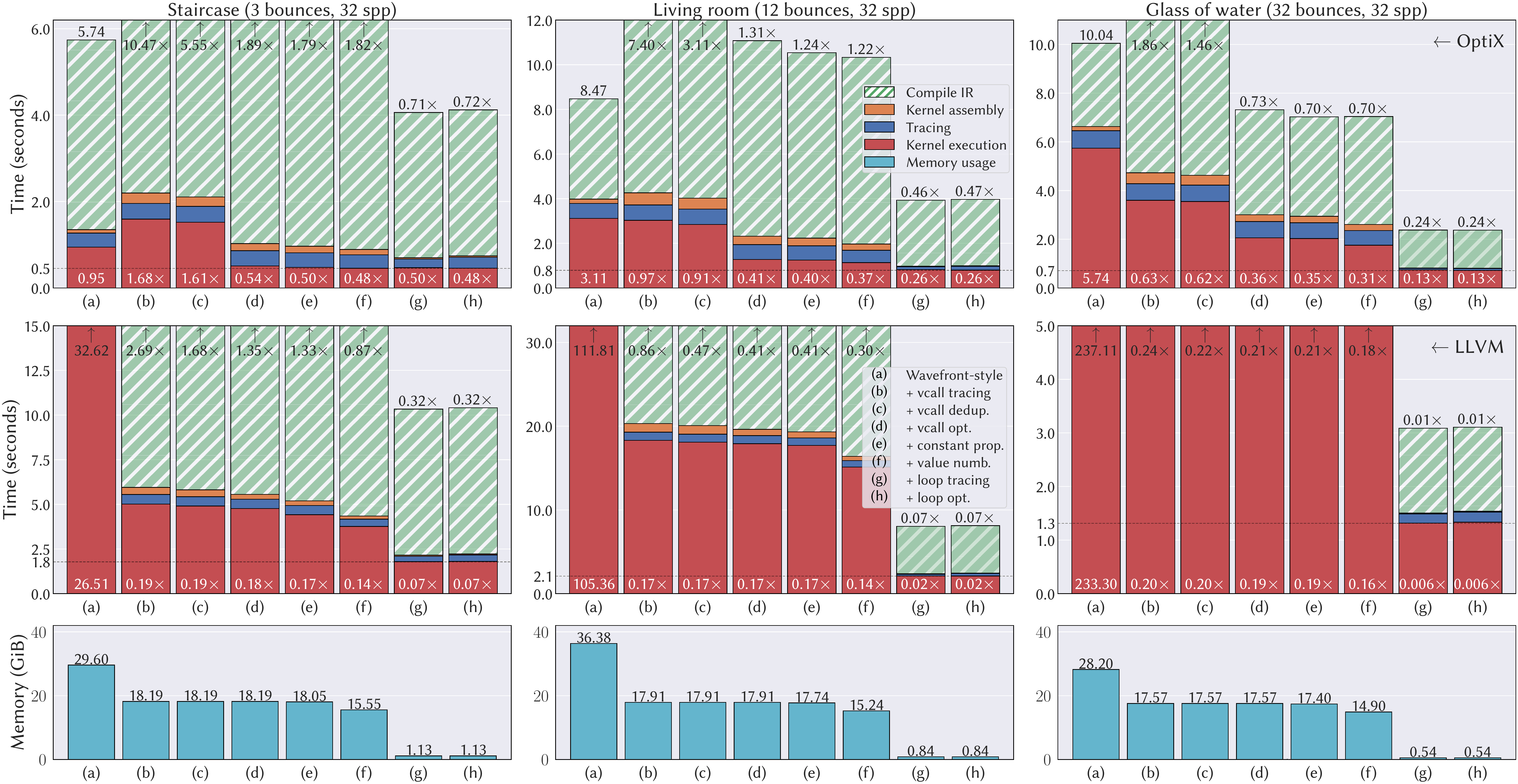}
    \vspace{-6.0mm}
    \caption{%
        \label{fig:benchmark-prb}%
        Reverse-mode differentiation benchmark of \emph{non-reparameterized}
        Path Replay Backpropagation analogous to
        Figure~\ref{fig:benchmark-primal-features}. We differentiate the
        rendered output of three scenes with respect to albedos and
        albedo textures, analyzing performance and the effect of different
        optimizations in \drjit.
        The rows show OptiX and LLVM runtime and peak memory usage (identical).
        Stacked bars indicate the time spent on backend IR
        compilation (hatched), kernel assembly (orange) and tracing
        (blue) within \drjit, as well as kernel execution (red).
        The wavefront baseline \textbf{(a)} on the left ``unrolls''
        all use of loops and polymorphism into separate kernels that
        communicate through global memory.
        Going to the right, we successively enable
        \textbf{(b)} constant propagation,
        \textbf{(c)} compilation of polymorphism into subroutines,
        \textbf{(d)} subroutine deduplication,
        \textbf{(e)} optimization of polymorphic calls,
        \textbf{(f)} local value numbering,
        \mbox{\textbf{(g)} loop tracing, and
        \textbf{(h)} loop optimizations.}
    }
\end{figure*}
\begin{figure*}[t]
    \centering
    \includegraphics[width=\textwidth]{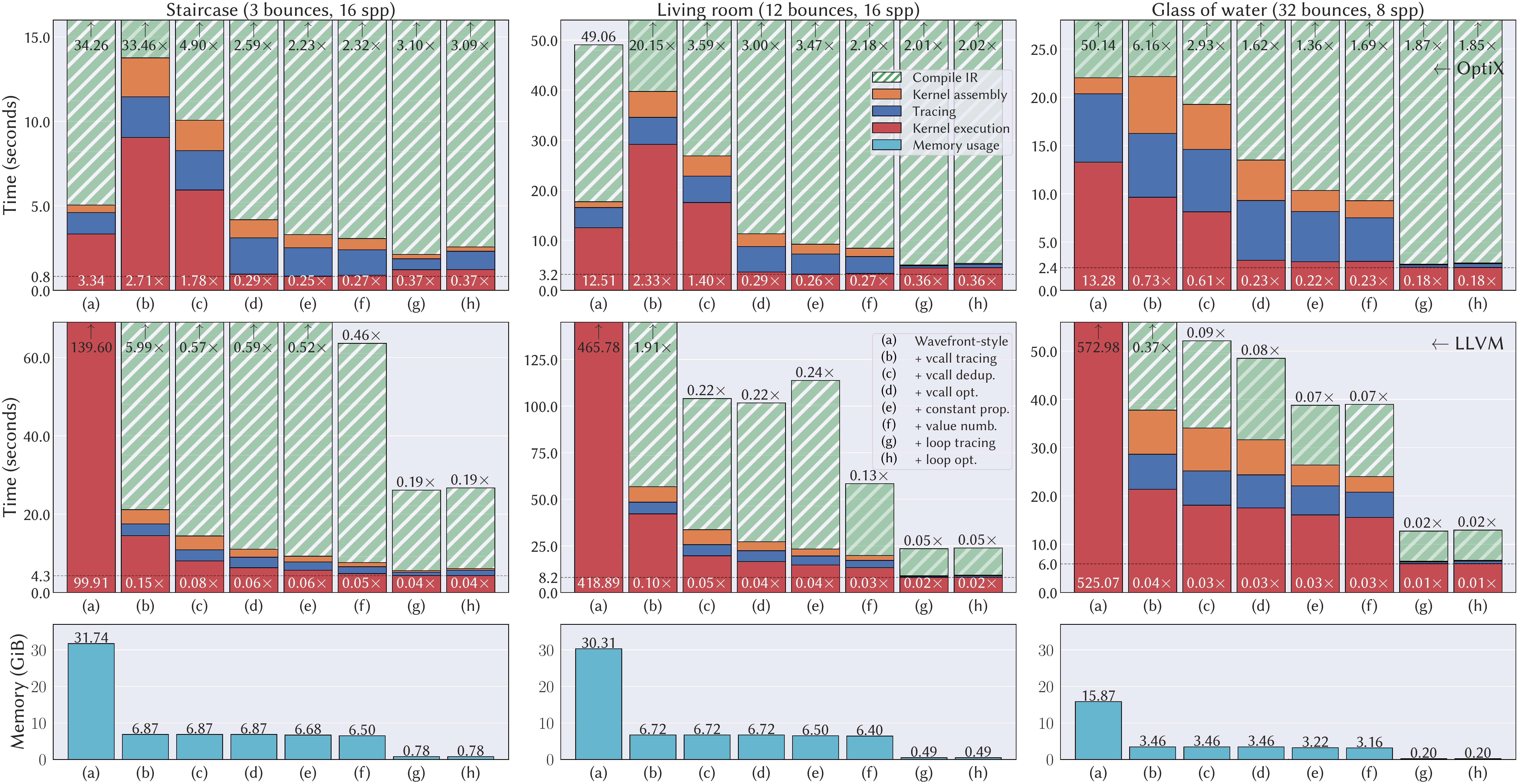}
    \vspace{-6.0mm}
    \caption{%
        \label{fig:benchmark-prb-reparam}%
        Reverse-mode differentiation benchmark of \emph{reparameterized} Path
        Replay Backpropagation analogous to
        Figures~\ref{fig:benchmark-primal-features}
        and~\ref{fig:benchmark-prb}. Please see their captions for details on
        the visualization, and Section~\ref{sec:results-2} for a discussion of
        this result.
    }
    \label{fig:benchmark_features_prb_reparam}
\end{figure*}

\subsection{AD tape surgery}
Implementations of PRB involve mysterious-looking steps like
\begin{minted}[fontsize=\footnotesize]{python}
Lr *= bsdf_value_diff / bsdf_value
\end{minted}
\noindent where the \texttt{bsdf\_value} and
\texttt{bsdf\_value\_diff} variables have the \emph{same value}. What purpose
do they serve?

In this example, \texttt{Lr} stores the reflected radiance obtained using a
\emph{detached} BSDF sampling strategy~\cite{Zeltner2021MonteCarlo}, which
means that the process of importance sampling was not differentiated. To
generate nonzero derivatives, the snippet above changes a factor within the
expression in \texttt{Lr} (specifically, \texttt{bsdf\_value}) so that it is
\emph{attached}, i.e., recomputed \emph{with} AD-based derivative tracking.

A problem with this approach is that it introduces unnecessary extra computation
in both primal and adjoint programs. We provide the
\mbox{\texttt{dr.replace\_grad(a, b)}} function that can be used
to accomplish this goal more efficiently. It
combines the primal value
of `\texttt{a}' with the AD trace of `\texttt{b}', which
changes the previous example to
\begin{minted}[fontsize=\footnotesize]{python}
Lr *= dr.replace_grad(1, bsdf_value_diff / bsdf_value)
\end{minted}
\vspace{1mm}
\noindent We also use this feature to
turn texture lookups performed using GPU
texture mapping units (TMUs)
into differentiable operations by attaching them to the
AD graph of a software-based lookup.

\subsection{Results}
\label{sec:results-2}
We now present results showcasing the combination of differentiation and
dynamic compilation. We not pursue complex application scenarios;
our focus is purely on the structure of the
computation and the performance of the system on such differential workloads.

\paragraph{Reverse-mode differentiation benchmark.} 
Using the same set of scenes as before (\emph{staircase}, \emph{living room},
\emph{glass of water}), we now use PRB to differentiate albedo values (scalar,
textured) and emitters like the environment map in the \emph{living room}
scene. The information in Figure~\ref{fig:benchmark-prb} only reflects the
differential portion of a gradient step (i.e., phases \#2 and \#3 of the
partition shown in Figure~\ref{fig:prb-structure}).

Many of the observations mirror our previous discussion of primal rendering in
section~\ref{sec:results-1}. On the CPU, the benefit of megakernel-style
evaluation continues to be dramatic, with speedups reaching $\sim\!166\,\times$
compared to the baseline ($\sim\!142\,\times$ in the primal benchmark).

One major change compared to the primal setting is that the computation must
now evaluate the reverse-mode derivative of polymorphic calls. Consider the
derivative of a function like BSDF evaluation that takes a large intersection
record as input. Its derivative has even more inputs, and it additionally
returns an output derivative for each primal input argument. Detecting and
removing the resulting redundancies using global dead code elimination,
constant propagation, and value numbering has a pronounced effect, with GPU
speedups from this alone reaching $3.3\,\times$ \mbox{in the
\emph{staircase} scene.}

The differential megakernels enabled by the methods presented in this article
generally achieve the lowest tracing and kernel assembly time,
lowest compilation time, and lowest runtime besides using a minimal amount of
GPU memory. This last point becomes relevant when optimizing large scene
representations (e.g. 3D volumes) that consume most of the available device
memory.

\paragraph{Reparameterized PRB benchmark.}
Figure~\ref{fig:benchmark-prb-reparam} repeats this experiment once more, using
\emph{reparameterized PRB}. The introduction of reparameterizations and steps
needed to evaluate the derivatives of adjacent vertices
(Section~\ref{label:fine-grained}) now leads to very large programs compared to
primal rendering or non-reparameterized methods.

This time, we optimize the vertex positions of all scene
geometry\footnote{Derivatives will be biased in the \emph{glass of water} scene
containing dielectric objects. How to reparameterize through perfectly specular
interfaces is an open research problem.}. Again, this experiment produces no
surprises and shows the effectiveness of the various optimizations. An
interesting contrast between Figures~\ref{fig:benchmark-prb}
and~\ref{fig:benchmark-prb-reparam} is the staircase-like sequence of bars in
the latter, which clearly shows the benefit that each of the separate steps
makes in both CUDA and especially the LLVM backend (which originally had a
relatively flat profile in Figure~\ref{fig:benchmark-prb}). Once more,
polymorphism-related optimizations are very important, with dead code
elimination, value numbering, and constant propagation now producing a
$6.6\times$ speedup on the OptiX backend.

The improvements of the loop state optimizations on runtime performance are
relatively modest (runtime improvements of a few percent in
Figure~\ref{fig:benchmark-prb}) compared to other
steps. We believe that this optimization may be more effective in
methods using \emph{attached} Monte Carlo
sampling~\cite{Zeltner2021MonteCarlo,Vicini2021}), where large sets of
derivatives must be exchanged between loop iterations.

\begin{figure}[t]
    \centering
    \includegraphics[width=\columnwidth]{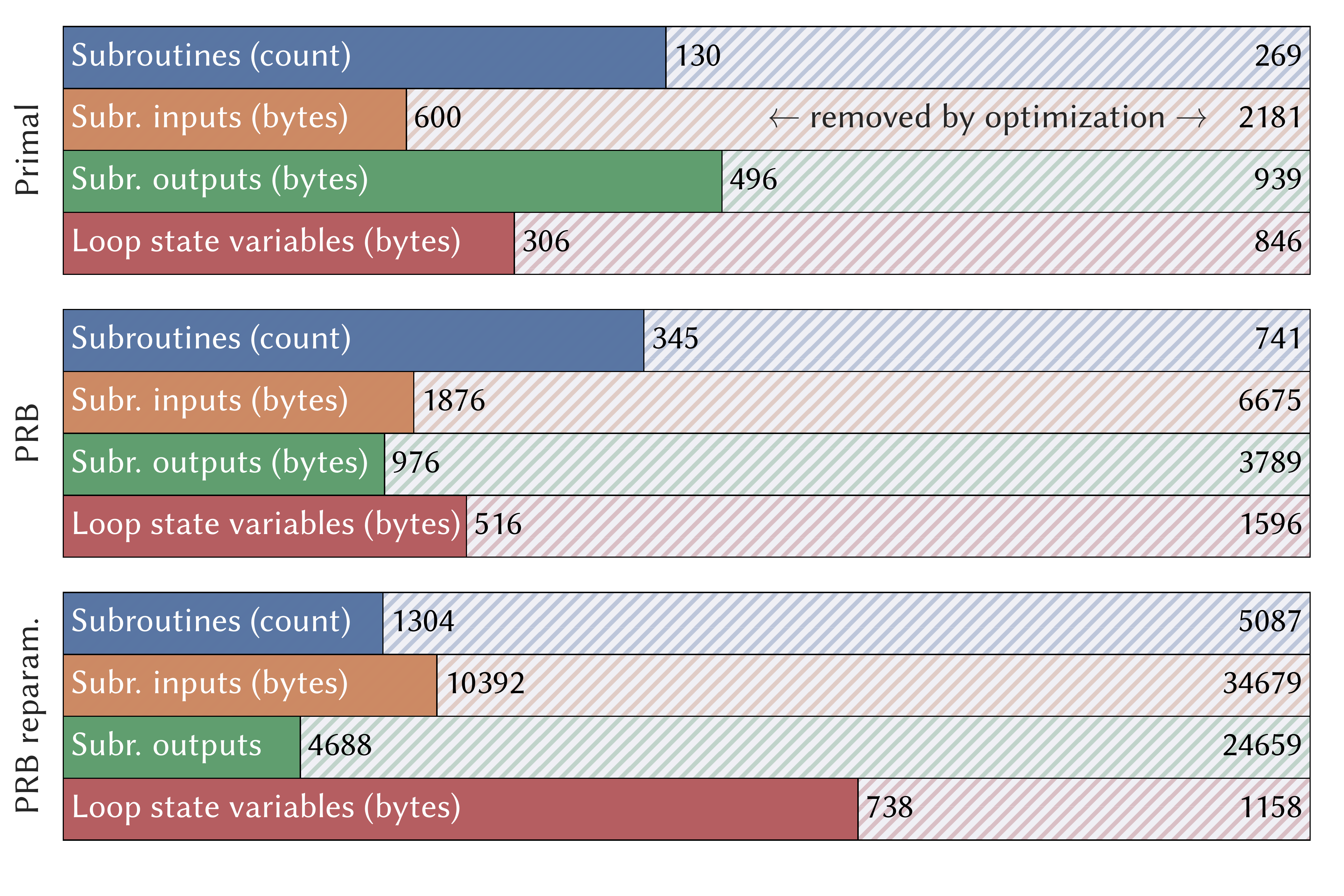}
    \vspace{-8mm}
    \caption{%
    \label{fig:measure_vcall_loop_opt_primal}%
        Visualization of the total number of generated kernel subroutines, and the
        amount of data exchanged through function inputs, outputs, and loop state
        variables. The stated amounts are sums across the three benchmark scenes.
        Dashed bars indicate the proportion that is removed by the presented
        optimizations. The three rows showcase the differences in behavior when
        compiling primal rendering, PRB, \mbox{and reparameterized PRB methods.}
}
\end{figure}

\vspace{-1mm}
\paragraph{Size reductions and size increases.}
Figure~\ref{fig:measure_vcall_loop_opt_primal} provides another view of the
effectiveness of optimizations besides compilation and runtime performance. It
shows that the majority of function arguments and return values are removed
regardless of the application. In reverse-mode (reparameterized) PRB, the
number of removed function outputs goes up significantly, as many computed
derivatives have no effect. Reparameterized PRB carries a large amount of loop
state to analyze the interaction between three adjacent path vertices; the
implementation of this method was carefully optimized, which explains the
smaller effect in the last row.
\addtolength{\belowcaptionskip}{-2mm}
\addtolength{\belowcaptionskip}{+2mm}

Table~\ref{table:kernel_sizes} gives an impression of how differentiation
increases the size of a rendering algorithm, as measured in the number of
compiled IR operations. The data indicates a relatively stable growth of
\mbox{$3\text{--}4\times$} when going from primal path tracing to
(differential) path replay. This is expected: besides steps for
differentiation, the method runs two separate rendering passes. When
reparameterizations are added on top, a different behavior emerges: wavefront
evaluation produces significant ($>30\times$) growth in the operation count,
which our optimizations then stabilize to a factor of $7\text{--}8\times$.

\paragraph{Differentiating existing methods}
\drjit can also differentiate standard methods that have not
been designed for this. Our last experiment demonstrates this using the
builtin Mitsuba (volumetric) path tracer and contrasts its behavior to PRB. The path tracer
contains a loop that \drjit handles using a checkpoint per scattering event
(Section~\ref{sec:tracing-dyn}). Differentiable polymorphism can still be used
within each iteration, reducing memory usage and communication costs.
Columns \textbf{(a)} and \textbf{(b)} show the performance of such an approach.

Differentiating with checkpoints is very memory-intensive but has the advantage
that only two Monte Carlo integration phases are needed in contrast to PRB's
three phases. Given a large supply of high-bandwidth memory (e.g. on the GPU),
checkpointing can yield acceptable performance. The runtime cost for PRB's
two-stage approach is initially higher in \textbf{(c)} (wavefront loop with
polymorphism) but outpaces checkpointing once compiled to a megakernel in
\textbf{(d)}. (These observations are consistent with the benchmarks reported
in the original paper~\cite{Vicini2021}.) Communication costs cause checkpointing to become less
competitive when the differentiated computation involves long-running loops, as shown in the
right example that differentiates a rendering with heterogeneous
subsurface scattering and a maximum path length of 128 interactions.

\section{Conclusion}

\drjit is a specialized compiler for physically based
differentiable rendering algorithms, whose unique set of constraints
makes their implementation using traditional means near-impossible. \drjit can
trace large object-oriented codebases with polymorphic indirections, while
providing fine-grained control over the differentiation process that is needed to exploit
\mbox{physical and mathematical symmetries.}

\makeatletter
\let\ftype@table\ftype@figure
\makeatother
\begin{figure}[t]
    \centering
    \includegraphics[width=.99\columnwidth]{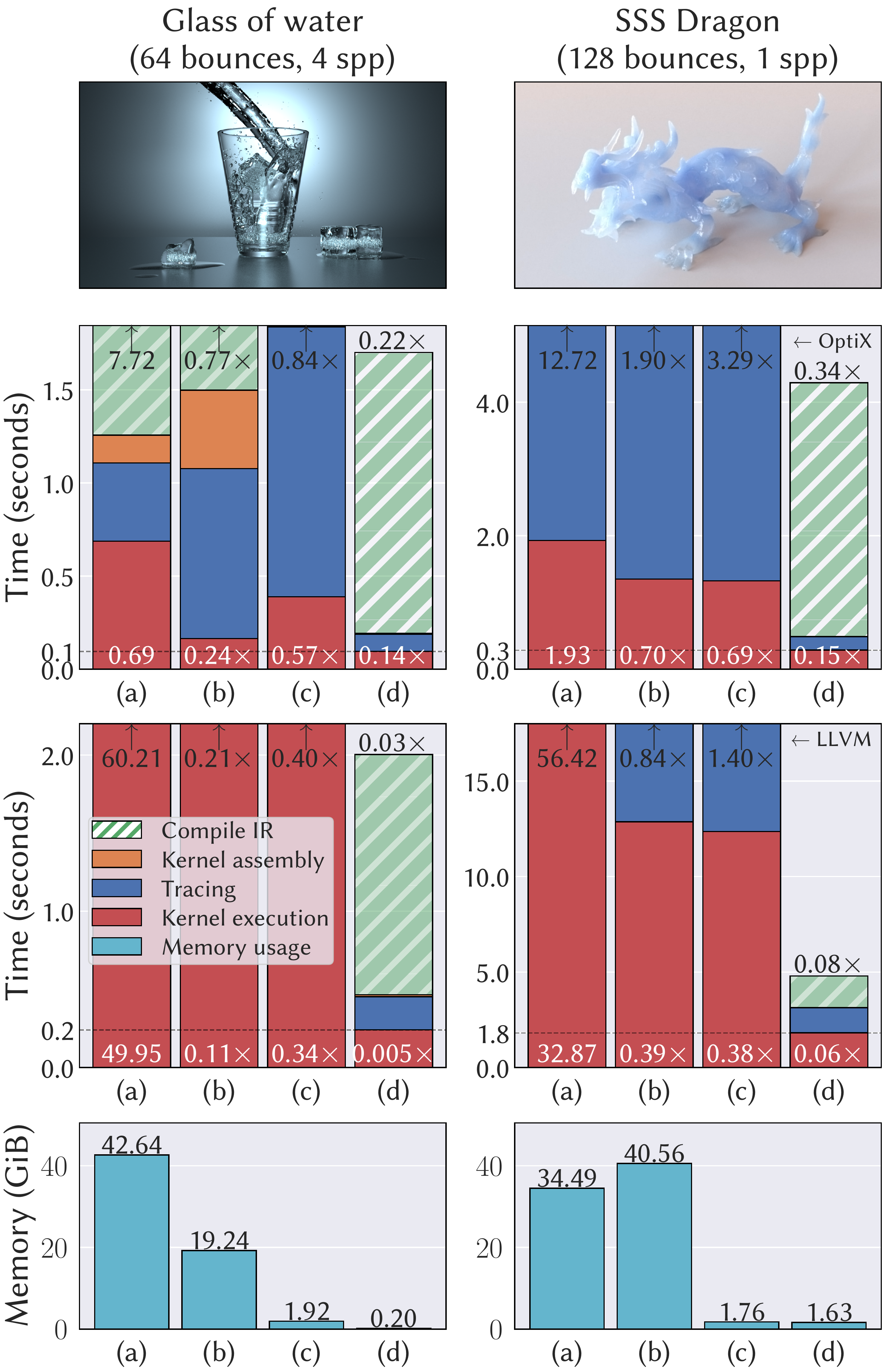}
    \vspace{-3mm}
    \caption{
        Standard methods like path tracing can be differentiated within \drjit
        but require checkpointing. This figure compares the behavior of such
        a memory-intensive approach to a specialized PBDR method.
        \textbf{(a)} Reverse-mode differentiation of a standard path tracer in 
        wavefront mode.
        \textbf{(b)} Path tracer with optimized polymorphic calls.
        \textbf{(c)} Wavefront PRB with optimized polymorphic calls.
        \textbf{(d)} \mbox{PRB compiled to a megakernel.}
    }
    \label{fig:benchmark_adjoint}
\end{figure}
\begin{table}[t!]
    \centering
    \caption{%
        \label{table:kernel_sizes}%
        Kernel sizes in thousands of IR operations averaged over the
        three benchmark scenes. This table provides numbers for primal rendering,
        reverse-mode (reparameterized) PRB, and the
        ratio relative to the primal column.
    }
    \vspace{-1mm}
    \begin{tabular}{l r r r r r}
        \toprule
                       & Primal      & PRB &   (ratio)      & Repa. PRB & (ratio) \\ 
        \midrule
        wavefront      & $188.94$ & $649.25$ & $\times\, 3.44$ & $6012.45$ & $\times\, 31.82$ \\
        + vcall rec.   & $216.28$ & $969.53$ & $\times\, 4.48$ & $6885.09$ & $\times\, 31.83$ \\
        + vcall dedup. & $141.21$ & $654.49$ & $\times\, 4.63$ & $1042.37$ & $\times\,  7.38$ \\
        + vcall opt.   & $138.74$ & $495.65$ & $\times\, 3.57$ & $1103.82$ & $\times\,  7.96$ \\
        + const prop.  & $127.13$ & $448.29$ & $\times\, 3.53$ & $ 907.36$ & $\times\,  7.14$ \\
        + value numb.  & $ 98.55$ & $334.59$ & $\times\, 3.39$ & $ 800.07$ & $\times\,  8.12$ \\
        + loop rec.    & $ 10.50$ & $ 36.45$ & $\times\, 3.47$ & $  78.77$ & $\times\,  7.50$ \\
        + loop opt.    & $ 10.49$ & $ 36.19$ & $\times\, 3.45$ & $  78.69$ & $\times\,  7.50$ \\
        \bottomrule
    \end{tabular}
\end{table}
Its combination of tracing at JIT and AD levels is harmonious: by tracking the
flow of primal and differential quantities globally, \drjit can
specialize algorithms to the problem at hand, while discarding redundant
computation. Its wavefront and megakernel implementations achieve excellent
\mbox{performance and run on diverse hardware.}

\drjit is not just meant to reproduce existing methods, but to provide a
foundation for future research. We hope that it will
lower the barrier to entry and enable new discoveries that push the boundaries
of physically based differentiable rendering.

\section{Acknowledgments}
The authors thank David Hart and the NVIDIA OptiX team for
tracking down a performance regression. Matt Pharr provided helpful
feedback on an early draft. This work was supported by the Swiss
National Science Foundation (SNSF) as part of grant 200021\_184629.

\bibliographystyle{ACM-Reference-Format}
\bibliography{paper.bib}
\appendix
\addtolength{\belowcaptionskip}{-4mm}
\begin{figure}[t]
    \centering
    \includegraphics[width=\columnwidth]{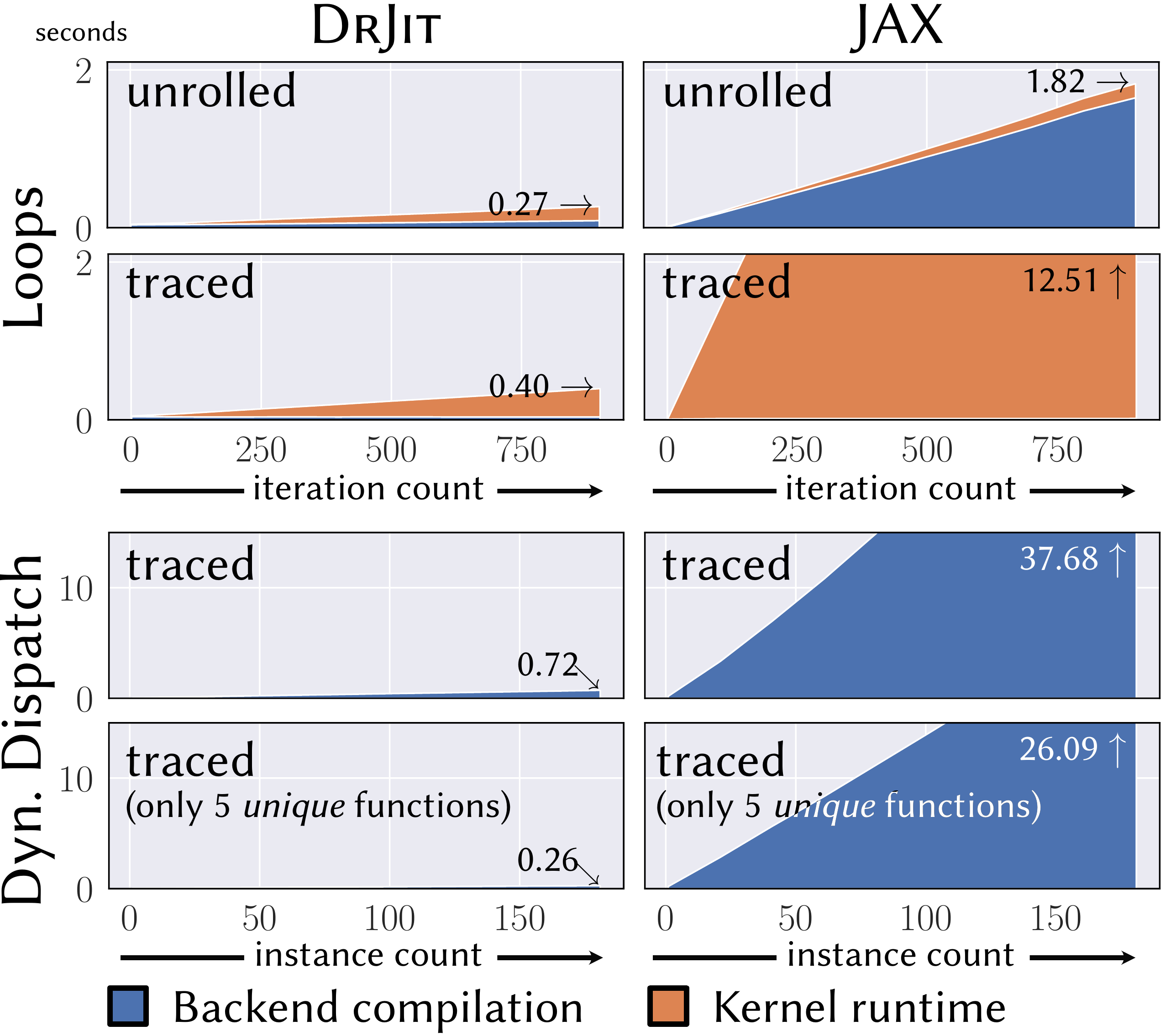}
    \caption{%
    \label{fig:jax-vs-drjit}%
Micro-benchmark comparing the tracing features of \drjit and JAX.
The top half benchmarks a simple loop with an update of the form
\texttt{x=(x+1)$\wedge$x} (\texttt{x} is an 1D array of $10^9$ 32-bit integers)
with loop counts increasing from 1 to 1000 on the horizontal axis and combined
compilation and runtime in seconds on the vertical axis.
The bottom half benchmarks dynamic dispatch to an increasingly large set of functions $f_1, f_2,
\ldots$ implementing successively better approximations of the sine power series, i.e., $f_i=\sum_{k=0}^i
-1^k / (2k+1)!\ x^{2k+1}$.
    In the last row, the functions $f_i$ internally take \texttt{i} modulo 5,
    which means that there are only 5 unique functions. This in principle
    provides an opportunity to greatly reduce the size of the program.
}
\end{figure}
\addtolength{\belowcaptionskip}{+4mm}
\section{Relationship to array programming}
\label{sec:appendix}
This section presents microbenchmarks comparing \drjit to a representative
array programming framework, specifically JAX~\cite{Jax}. JAX is powerful a
frontend to the XLA compiler~\cite{XLA}, which analyzes the graph structure of
the desired computation, \emph{fusing} sequences operations as part of a
heuristically guided clustering process. This involves tensorial optimizations
like merging sequences of matrix-vector multiplications into matrix-matrix
multiplications and selecting among the thousands of vendor-tuned kernels
included in libraries like NVIDIA's cuDNN~\cite{cuDNN}. Steps that are not
handled by an existing kernel require just-in-time compilation. JAX and XLA
provide functionality to handle such custom computation including loop and
dynamic function dispatch tracing resembling that of \drjit. We initially
implemented a small renderer using these primitives, but found that compilation
timed out on nontrivial examples. Figure~\ref{fig:jax-vs-drjit} benchmarks XLA
tracing primitives to better understand these issues.

When unrolling loops, we observe significantly increased backend compilation
time in JAX. Tracing loops via \texttt{jax.lax.fori\_loop()} addresses this,
but the generated code curiously commits loop state to memory at every
iteration, which increases runtime cost. JAX also provides
\texttt{jax.lax.switch()}, which produces a \mbox{\texttt{switch\{\}}-like}
statement that effectively merges all functions into the body of the generated
kernel. This produces a very large IR representation that exacerbates the cost
of steps like register allocation. \drjit instead generates indirect calls to
subroutines that admit separate compilation. In the context of rendering,
dynamic dispatch often targets functions that later turn out to be identical,
which \drjit exploits to reduce backend compilation time to a constant. XLA
detects and exploits these redundancies as well, albeit with a runtime cost
that grows with the size of the input program.

\section{Dynamic dispatch to closures}
\label{sec:closures}
Instance attributes present a subtle but important challenge that must be
handled while tracing methods. Consider the following rudimentary
implementation of a textured Phong BRDF:
\begin{minted}[fontsize=\footnotesize,escapeinside=||,mathescape=true]{python}
class Phong(mitsuba.BSDF):
    def __init__(self, albedo: TensorXf, exponent: Float):
        self.albedo = albedo      # Bitmap data (TensorXf)
        self.exponent = exponent  # Specularity (scalar Float)

    def eval(self, si: SurfaceInteraction3f, wo: Vector3f):
        # Convert UV into texel coordinates and perform a lookup
        resolution = Vector2u(self.albedo.shape)
        pos = dr.min(Vector2u(si.uv * resolution), resolution - 1)
        albedo = self.albedo[pos.y, pos.x]
        # Evaluate reflected direction and BRDF terms
        r = Vector3f(-si.wi.x, -si.wi.y, si.wi.z)
        return albedo * dr.inv_pi + dr.dot(r, wo) ** self.exponent
\end{minted}
\noindent Evaluation accesses two attributes: \texttt{albedo}, which refers to
a region in device memory, and a scalar \texttt{exponent} controlling
specularity.

Tracing \texttt{eval()} must eventually produce a LLVM or PTX subroutine, whose
signature is consistent with other implementations of this interface. It is
logical that \emph{explicit} function arguments like \texttt{si} and
\texttt{wo} will be part of this interface, but how should \emph{implicit}
dependencies like \texttt{self.albedo} and \texttt{self.exponent} be handled?
Each instance may reference different numbers and types of attributes, hence
they cannot easily be turned into explicit parameters. It is also important to
realize that ``\texttt{self}'' is purely a Python construct at this point that
lacks a device-specific representation. A na\"ive solution would be to embed
the raw attribute values in the generated IR:
\begin{minted}[fontsize=\footnotesize,escapeinside=||,mathescape=true]{python}
def eval_impl_001(si_uv_x: float, si_uv_y: float, ...) -> float:
    albedo   = 0xfffe2f00 # Pointer to device memory
    exponent = 10         # Hardcoded constant
    # ... implementation (this would be LLVM/PTX IR in practice) ...
\end{minted}
\noindent This is undesirable because it would break function deduplication,
causing every \texttt{Phong} instance to produce a unique subroutine.

Instead of merely tracing functions, the system must handle \emph{closures},
which refers to pairings of functions with data from a surrounding environment.
Tracing and kernel assembly should then separate code from data so that
optimizations remain effective. \drjit transparently performs this optimization
by detecting implicit variable dependencies while tracing function calls. Their
contents are written into a contiguous array along with an auxiliary offset
array that associates a closure data block with each instance.
\begin{center}
    \includegraphics[width=.9\columnwidth]{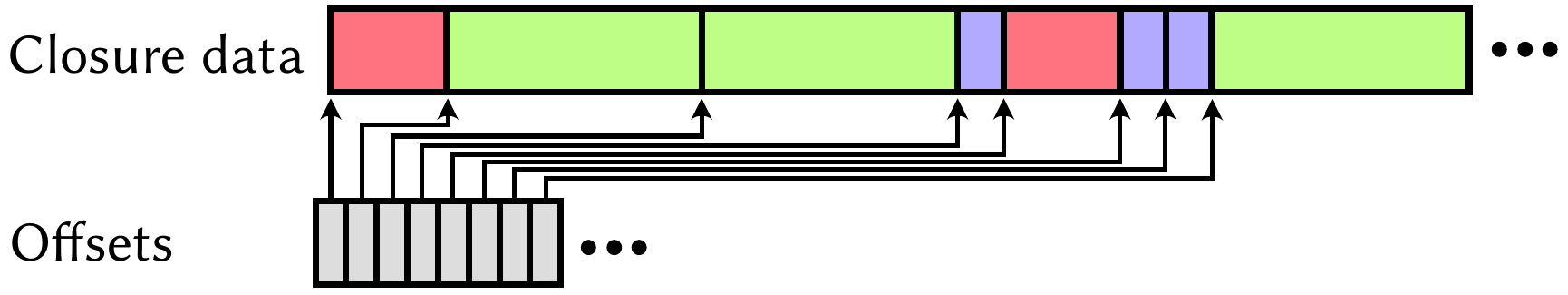}
\end{center}
\drjit launches a builtin asynchronous kernel that collects dependencies from
various locations in device memory to build this consolidated data structure.
The total amount of device memory needed for them is small---usually on the
order \mbox{of 10--100 KiB per method call}\footnote{Large device arrays like
textures or volumes are accessed \emph{indirectly}. The closure data array
records a pointer to them rather than the data itself, while scalars are
directly copied to avoid an extra indirection.}.

\end{document}